\documentclass[
  aps,  
  prx,  
  superscriptaddress,  
  raggedbottom,  
  floatfix,  
  twocolumn,  
  10pt,  
  longbibliography,  
  noeprint,  
]{revtex4-2}

\usepackage[T1]{fontenc}  
\usepackage[utf8]{inputenc}  

\usepackage[english]{babel}
\usepackage{amsmath}
\usepackage{amssymb}
\usepackage{graphicx}
\usepackage{extarrows}
\usepackage[hyperref]{xcolor}
\usepackage[pdfusetitle]{hyperref}
\usepackage{cleveref}
\usepackage{dsfont}
\usepackage{float}
\usepackage{braket}

\hypersetup{
  pdfsubject={Quantum Physics},  
  pdfauthor={Johannes Mögerle},
  colorlinks=true,  
  linkcolor={red!60!black},  
  citecolor={blue!80!black},  
  urlcolor={blue!95!black},  
}

\newcommand{\abs}[1]{\left| #1 \right|}
\newcommand{\hc}{\text{h.c.}}
\newcommand{\ketbra}[2]{\ket{#1} \! \bra{#2}}
\newcommand{\expvalue}[1]{\left\langle #1 \right\rangle}

\newcommand{\qty}[2]{\ensuremath{#1\,\mathrm{#2}}}
\newcommand{\unit}[1]{\ensuremath{\mathrm{#1}}}

\newcommand{\mzero}{}

\begin{document}
\title{Spin-1 Haldane phase in a chain of Rydberg atoms}

\author{J.~Mögerle}
\affiliation{Institute for Theoretical Physics III and Center for Integrated Quantum Science and Technology, University of Stuttgart, Pfaffenwaldring 57, 70569 Stuttgart, Germany}

\author{K.~Brechtelsbauer}
\affiliation{Institute for Theoretical Physics III and Center for Integrated Quantum Science and Technology, University of Stuttgart, Pfaffenwaldring 57, 70569 Stuttgart, Germany}

\author{A.~T.~Gea-Caballero}
\affiliation{Institute for Theoretical Physics III and Center for Integrated Quantum Science and Technology, University of Stuttgart, Pfaffenwaldring 57, 70569 Stuttgart, Germany}
\affiliation{Departamento de Física - CIOyN, Universidad de Murcia, Murcia E-30071, Spain}
\author{J.~Prior}
\affiliation{Departamento de Física - CIOyN, Universidad de Murcia, Murcia E-30071, Spain}

\author{G.~Emperauger}
\affiliation{Université Paris-Saclay, Institut d'Optique Graduate School, CNRS, Laboratoire Charles Fabry, 91127 Palaiseau Cedex, France}
\author{G.~Bornet}
\affiliation{Université Paris-Saclay, Institut d'Optique Graduate School, CNRS, Laboratoire Charles Fabry, 91127 Palaiseau Cedex, France}
\author{C.~Chen}
\affiliation{Université Paris-Saclay, Institut d'Optique Graduate School, CNRS, Laboratoire Charles Fabry, 91127 Palaiseau Cedex, France}
\author{T.~Lahaye}
\affiliation{Université Paris-Saclay, Institut d'Optique Graduate School, CNRS, Laboratoire Charles Fabry, 91127 Palaiseau Cedex, France}
\author{A.~Browaeys}
\affiliation{Université Paris-Saclay, Institut d'Optique Graduate School, CNRS, Laboratoire Charles Fabry, 91127 Palaiseau Cedex, France}

\author{H.~P.~Büchler}
\affiliation{Institute for Theoretical Physics III and Center for Integrated Quantum Science and Technology, University of Stuttgart, Pfaffenwaldring 57, 70569 Stuttgart, Germany}

\date{\today}

\begin{abstract}
  We present a protocol to implement a spin-1 chain in Rydberg systems using three Rydberg states close to a Förster resonance.
  In addition to dipole-dipole interactions, strong van der Waals interactions naturally appear due to the presence of the Förster resonance and give rise to a highly tunable Hamiltonian.
  The resulting phase diagram is studied using the infinite density-matrix renormalization group and reveals a highly robust Haldane phase -- a prime example of a symmetry protected topological phase.
  We find experimentally accessible parameters to probe the Haldane phase in current Rydberg systems, and demonstrate an efficient adiabatic preparation scheme.
  This paves the way to probe the remarkable properties of spin fractionalization in the Haldane phase.
\end{abstract}

\maketitle

\section{Introduction\label{sec:intro}}

Topological phases in quantum many-body systems exhibit intriguing properties like anyonic excitations, topological order, string order, edge states and fractionalization~\cite{wenZooQuantumtopologicalPhases2017}.
Many of these topological phases that are currently studied theoretically as well as experimentally, like the fractional quantum Hall state~\cite{stormerFractionalQuantumHall1999} and spin liquids~\cite{verresenPredictionToricCode2021a,semeghiniProbingTopologicalSpin2021}, occur in two or higher dimensional systems.
For one-dimensional systems, it was shown that such topologically ordered phases do not exist~\cite{chenClassificationGappedSymmetric2011}.
However, so-called symmetry protected topological (SPT) phases can appear in one dimension and still feature many similar properties.
Thus, such one-dimensional SPT phases can often be used to get a better understanding of these topological properties.
A prime example for this is the spin-1 Haldane phase~\cite{haldaneNonlinearFieldTheory1983,haldaneContinuumDynamics1Heisenberg1983} and its characteristic Affleck-Kennedy-Lieb-Tasaki (AKLT) state~\cite{affleckRigorousResultsValencebond1987}, which introduced the concept of Valence Bond Solids (VBS) and hidden order.
Its most intriguing property, however, is the appearance of fractionalized spins at the edges.
In this letter, we propose a scheme to realize such a spin-1 Haldane phase using Rydberg atoms and provide insights into how to detect the fractional spin states in an experiment.

In the past, there has been a large experimental effort to study the Haldane phase in quasi-one-dimensional chains of transition metal ions~\cite{renardHaldaneQuantumSpin2001},
and recently signatures of fractional edge states have been reported in nano-graphene~\cite{mishraObservationFractionalEdge2021}.
Additionally, most recently the AKLT state was constructed on a trapped-ion quantum processor~\cite{edmundsConstructingSpin1Haldane2024}.
On the other hand, there is a large effort in studying strongly correlated states of matter using quantum simulators~\cite{altmanQuantumSimulatorsArchitectures2021,grossQuantumSimulationsUltracold2017,monroeProgrammableQuantumSimulations2021}
and theoretical proposals for the Haldane phase have been made for cold atomic gases in optical lattices~\cite{dallatorreHiddenOrder1D2006,brennenDesigningSpin1Lattice2007}, ion traps~\cite{cohenProposalVerificationHaldane2014,senkoRealizationQuantumIntegerSpin2015}, and superconductor-semiconductor hybrids~\cite{baranSpin1HaldaneChains2024}.
Especially, Rydberg atoms have emerged as a highly promising platform for exploring strongly correlated quantum many-body systems~\cite{browaeysManyBodyPhysicsIndividuallyControlled2020}
due to their tunable interactions, including dipole-dipole and van der Waals interactions~\cite{weberTopologicallyProtectedEdge2018,deleseleucObservationSymmetryprotectedTopological2019,lienhardRealizationDensitydependentPeierls2020,sorensenFractionalQuantumHall2005,buchlerAtomicQuantumSimulator2005,duanControllingSpinExchange2003,semeghiniProbingTopologicalSpin2021}.
Additionally, the use of optical tweezers allows for precise control over individual atoms~\cite{miroshnychenkoPrecisionPreparationStrings2006,kimSituSingleatomArray2016,endresAtomatomAssemblyDefectfree2016,barredoAtomatomAssemblerDefectfree2016}.
Various quantum many-body phases have already been realized in such Rydberg systems in recent years, ranging from the one-dimensional Su-Schrieffer-Heeger (SSH) model~\cite{deleseleucObservationSymmetryprotectedTopological2019} to two-dimensional quantum Ising models~\cite{labuhnTunableTwodimensionalArrays2016}, and even signatures of spin liquids have been observed in Rydberg systems~\cite{semeghiniProbingTopologicalSpin2021}.
While most implemented models reduce each Rydberg atom to an effective two-level system, there also are theoretical proposals that utilize three levels in a \emph{V}-shaped configuration to implement two hard-core bosonic particles~\cite{weberExperimentallyAccessibleScheme2022}.
In the presence of Förster resonances~\cite{safinyaResonantRydbergAtomRydbergAtomCollisions1981,ryabtsevObservationStarkTunedForster2010}, equidistant three-level systems are realized, leading to additional dipole-dipole interactions and van der Waals terms~\cite{MScJohannesMoegerle, liuSupersoliditySimplexPhases2024a, homeierAntiferromagneticBosonic2024}.

Here, we present a detailed study on how these three Rydberg states, when close to a Förster resonance, can be used to implement an effective spin-1 system.
The typical dipole-dipole and van der Waals interactions of these Rydberg states can then be mapped to spin-1 operators, resulting in an effective spin-1 Hamiltonian.
By tuning the various parameters of the Rydberg system we can access different regimes in the phase diagram of the spin-1 model.
We study the different phases using the infinite density-matrix renormalization group (infinite DMRG) and demonstrate the appearance of the Haldane phase for experimentally realistic parameters.
Furthermore, we show that the ground state of the Haldane phase can be adiabatically prepared by starting from a product state and applying a staggered, effective magnetic field.
Signatures of the Haldane phase, such as the string order parameter and edge states, can be experimentally observed through site-resolved measurements of the local spins.
While some experimental signatures of the Haldane phase have already been observed in the topological equivalent SSH chain using qubit systems~\cite{deleseleucObservationSymmetryprotectedTopological2019,sompetRealisingSymmetryProtectedHaldane2022}, the Förster resonance enables us to implement a spin-1 chain and thus allow us to additionally study the fractionalization of the edge spin degrees of freedoms in the Haldane phase.

\section{System and Model\label{sec:model}}
\begin{figure}
  \centering
  \includegraphics{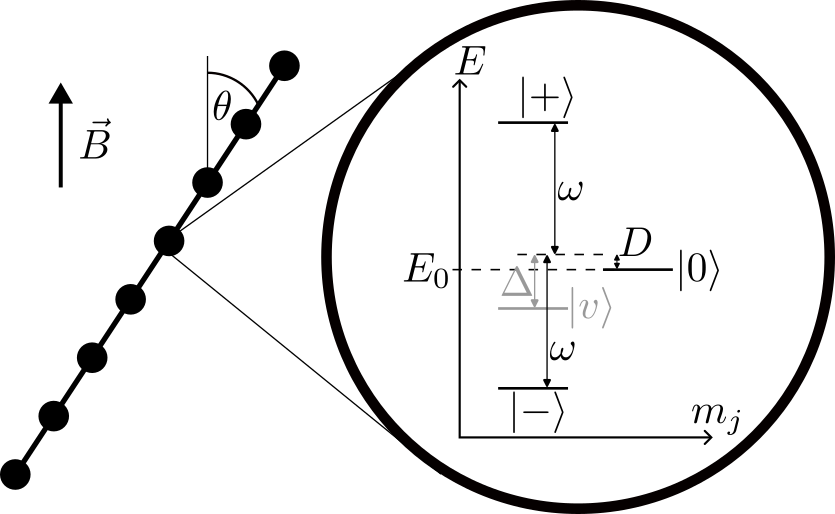}
  \caption{
    Sketch of the one-dimensional chain of Rydberg atoms aligned at an angle $\theta$ with respect to the magnetic field $\vec{B}$.
    In addition, a simplified level structure of the Rydberg atoms is shown.
    This includes the three main states $\ket{+}, \ket{0}, \ket{-}$, which are defined in \cref{eq:states}.
    Furthermore, an additional nearby state $\ket{v}$ is shown, which explains the dominant part of the van der Waals interactions.
    \label{fig:system}
  }
\end{figure}
\begin{figure}
  \centering
  \includegraphics{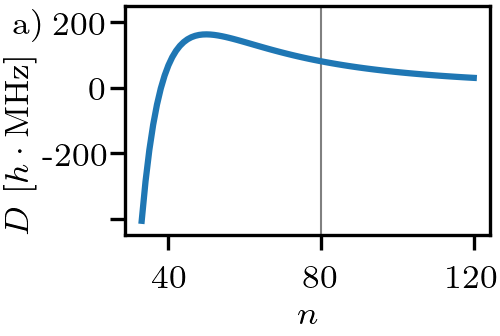}
  \includegraphics{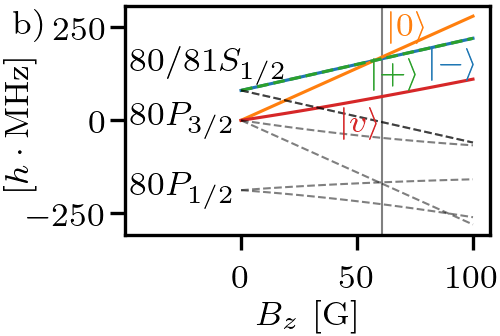}
  \includegraphics{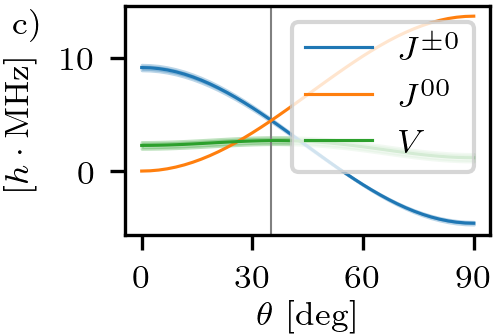}
  \includegraphics{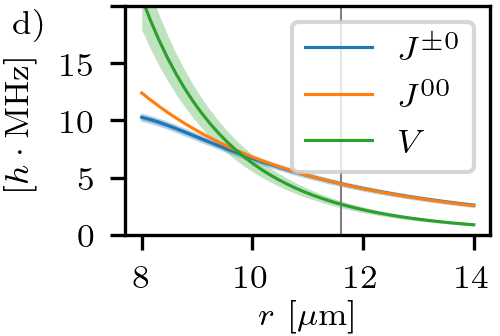}
  \caption{
    All energies and interactions in this figure are calculated for Rubidium via the \emph{pairinteraction} software~\protect{\cite{weberTutorialCalculationRydberg2017}}.
    a) The anisotropy energy $D = (E_+ + E_-)/2 - E_0$ (without any fields applied) for the states defined in \cref{eq:states}.
    b) The energy spectrum of the Rydberg Zeeman sublevels of interest in a rotating frame with respect to the energy $E_0$ of the state $\ket{0}$ at $B=0$.
    The $80S_{1/2}$ and $81S_{1/2}$ manifolds are shifted by $\pm \omega$ and thus lie on top of each other.
    c) and d) The interaction strengths $J^{\pm 0}$, $J^{00}$ and $V$ over c) the angle $\theta$ between the magnetic field and the interatomic axis and d) the interatomic distance $r$.
    For c) and d) we applied a magnetic field of $B = \qty{60.7}{G}$ and the distance (angle) was fixed at $r = \qty{11.6}{\mu m}$ ($\theta = \qty{35.1}{^\circ}$).
    The blue shaded area around the $J^{\pm 0}$ line indicates the difference of the $J^{+0}$ and $J^{-0}$ interaction strengths
    and the green shaded area around the $V$ line indicates the difference of the $V^\text{diag}$ and $V^\text{offd}$ interaction strengths.
    The gray vertical lines in the individual plots indicate the values used in \cref{sec:exp-proposal}.
  }
  \label{fig:pairinteraction}
\end{figure}
Our system, which is sketched in \cref{fig:system}, consists of a one-dimensional chain of Rydberg atoms.
From the Rydberg manifold of each atom, we choose three states, $\ket{+}, \ket{0}, \ket{-}$, as our states of interest.
These states are given by
\begin{align}
  \ket{+} &= \ket{(n + 1) S_{1/2}, m_j = +1/2}
  \,, \nonumber \\
  \ket{0} &= \ket{n P_{3/2}, m_j = +3/2}
  \,, \nonumber \\
  \ket{-} &= \ket{n S_{1/2}, m_j = +1/2}
  \,,
  \label{eq:states}
\end{align}
which are quantized along the magnetic field $\vec{B} = B \vec{e_z}$.
By tuning the magnetic field $B$, we make sure that in the rotating wave frame these states are energetically well separated from the other Rydberg states (see \cref{fig:pairinteraction}~b)).

Note that we label our Rydberg states like spin-1 states, since the three Rydberg states can be used to effectively describe a spin-1 system.
In the remainder of this section we show that the many-body Hamiltonian of this Rydberg system for a proper choice of the experimental parameters can effectively describe the following spin-1 Hamiltonian
\begin{align}
  H &=
  \frac{1}{2} \sum_{i \neq j} \frac{J^{xy}}{r_{ij}^3} \left( S^x_i S^x_j +S^y_i S^y_j \right)
  + D \sum_{i} (S^z_i)^2
  \nonumber \\
  &\quad + \frac{1}{4} \sum_{i \neq j} \frac{V}{r_{ij}^6}
  \Bigl[ (S^+_i)^2 (S^-_j)^2 + (S^z_i S^z_j - 1) S^z_i S^z_j \Bigr]
  \label{eq:many-body-H}
  \,,
\end{align}
where the $S^{x,y,z,\pm}$ are spin-1 operators and are explicitly defined in \cref{app:rydberg-to-spin}.
$J^{xy}$ is the dipole-dipole interaction strength, $D$ is the anisotropy energy, and $V$ is the van der Waals interaction strength.
Note that throughout this paper, we express all energies as frequencies in units of $h \cdot \unit{MHz}$, where $h$ is the Planck constant.

While the Haldane phase was originally introduced for the $SO(3)$ symmetric Heisenberg model~\cite{haldaneNonlinearFieldTheory1983,haldaneContinuumDynamics1Heisenberg1983}, it was later shown that even the dihedral symmetry $D_2$ ($= \mathbb{Z}_2 \times \mathbb{Z}_2$) or time reversal symmetry can protect the Haldane phase~\cite{chenClassificationGappedSymmetric2011,pollmannEntanglementSpectrumTopological2010}.
The $D_2$ symmetry is a subgroup of the $SO(3)$ symmetry and consists of all $\pi$ rotations around the $x$-, $y$- and $z$-axis.
The Hamiltonian~\eqref{eq:many-body-H} respects this dihedral symmetry $D_2$ as well as time reversal symmetry.

A famous example of the Haldane phase is the AKLT model~\cite{affleckRigorousResultsValencebond1987}, whose ground state can be described by considering each spin-1 particle as two symmetrized spin-$1/2$ particles.
Then two adjacent spin-$1/2$ particles (from different spin-1 particles) form a singlet state (Valence Bond).
By this construction, the AKLT state on an open chain has two uncoupled spin-$1/2$ degrees of freedom at the edges.
These fractionalized edge states are not a specific features of the AKLT model, but a general property of the Haldane phase as long as one of the $D_2$ or time reversal symmetry is preserved~\cite{pollmannEntanglementSpectrumTopological2010}.
The emergence of these fractionalized edge states can also be related to the fractionalization of the protecting symmetries~\cite{pollmannDetectionSymmetryProtected2012}.
Since the bulk must be invariant under the protecting symmetry, e.g. the $SO(3)$ symmetry, the symmetry splits into a tensor product acting only on the left and right edge.
The resulting projective symmetry is the $SU(2)$ group, which acts on the spin-$1/2$ degrees of freedom on the left and right edge.

It was shown that the spin-1 XXZ-D model [this corresponds to the first line of \cref{eq:many-body-H} and an additional coupling $J^z S^z_i S^z_j$] for only nearest neighbor interactions does feature the Haldane phase, but only for $J^z > 0$~\cite{chenGroundstatePhaseDiagram2003}.
By adding long range interactions one can stabilize the Haldane phase for the antiferromagnetic spin-1 XXZ model~\cite{gongKaleidoscopeQuantumPhases2016} also for $J^z =0$.
However, the signatures in the region $J^z = 0$ are rather weak, and it is not clear, whether the Haldane phase could be measured in a finite size experiment.
Here, we demonstrate that natural strong van der Waals terms appear due to the Förster resonance.
These additional terms can further stabilize the Haldane phase and make it more accessible in a real experiment.
We emphasize that the van der Waals term consists of multiple parts:
One term, which is proportional to $S^z_i S^z_j$, and thus drives the model closer to the Heisenberg point, possibly resulting in a more stable Haldane phase;
and a second term, which is proportional to $(S^+_i)^2 (S^-_j)^2 + (S^-_i)^2 (S^+_j)^2$.
This second term is also part of the AKLT Hamiltonian~\cite{affleckRigorousResultsValencebond1987}, and thus can stabilize the Haldane phase.

In the remainder of this section we motivate and explain the three terms $D$, $J^{xy}$ and $V$ in more detail on the level of the Rydberg states.
A mapping of the Rydberg interactions of the three states $\ket{+}, \ket{0}, \ket{-}$ to spin-1 operators used in \cref{eq:many-body-H} can be found in \cref{app:rydberg-to-spin}.

\subsection*{Anisotropy energy}
Ignoring a global energy shift $E_0$, we have two energies in this system: $\omega = (E_+ - E_-)/2$ and $D = (E_+ + E_-)/2 - E_0$ (see \cref{fig:system}).
For the states we are interested in, the energy difference $\omega$ is much larger than the anisotropy energy $D$.
This allows us to treat the system as a spin-1 system, where the $z$-component of the total magnetization $M_\mathrm{tot} = \expvalue{\sum_i S^z_i}$ is conserved.

To achieve this desired energy structure, we use the Rydberg states proposed in \cref{eq:states}.
In \cref{fig:pairinteraction}~a), one can see that for different values of $n$ we are indeed close to so-called Förster resonances, where the anisotropy energy $D$ is small compared to the energy difference $\omega$, which for $n=80$ is $\omega/h \approx \qty{7}{GHz}$.
Often the Förster resonance at $n = 37$ is used, because here the anisotropy energy is very small and negative, which allows one to tune the anisotropy energy $D$ to zero by applying an electric field.
In the following we focus on the case $n = 80$, where for the Zeeman sublevels specified in \cref{eq:states} one can tune $D$ closer to zero by applying a magnetic field.
However, the results are qualitatively the same for other choices of $n$.
By applying a magnetic field $B$ we induce a Zeeman shift on all states, which splits the different magnetic sublevels and thus tunes the anisotropy energy $D$.
This can be seen in \cref{fig:pairinteraction}~b), where we plotted the $80S_{1/2}$ and $81S_{1/2}$ in a rotating frame (shifted by the energy $\pm \omega$) together with the $80P_{3/2}$ and $80P_{1/2}$ manifolds.
For magnetic fields $B \approx \qty{60}{G}$ the three states $\ket{+}$, $\ket{0}$ and $\ket{-}$ are nearby degenerate and well separated from the other Zeeman sublevels, which are plotted as gray dashed lines.

The choice of the quantum numbers $l$, $j$ and $m_j$ above is partly motivated by the Förster resonance and partly motivated by constructing exactly the following Rydberg interactions that map to typical spin-1 interactions.

\subsection*{Dipole-dipole interactions}
The allowed dipole-dipole interactions in our Rydberg system are \\
$\ket{+0} \xrightarrow{d_i^+d_j^-} \ket{0+}$
with strength
\begin{equation*}
  J^{+0} = \frac{3 \cos^2 \theta - 1}{2 r^3} \abs{\bra{0} d^+ \ket{+}}^2
  \,,
\end{equation*}
$\ket{-0} \xrightarrow{d_i^+d_j^-} \ket{0-}$
with strength
\begin{equation*}
  J^{-0} = \frac{3 \cos^2 \theta - 1}{2 r^3} \abs{\bra{0} d^+ \ket{-}}^2
  \,,
\end{equation*}
and $\ket{+-} \xrightarrow{d_i^+d_j^+} \ket{00}$ and $\ket{-+} \xrightarrow{d_i^+d_j^+} \ket{00}$
with strength
\begin{equation*}
  J^{00} = \frac{- 3 \sin^2 \theta}{2 r^3} \bra{0} d^+ \ket{-} \bra{0} d^+ \ket{+}
\end{equation*}
as well as all hermitian conjugates.
Here, the $d^{\pm}$ are the dipole operators of the Rydberg states, acting on the total angular momentum $j$ and $m_j$.

For large principal quantum numbers $n$, the dipole moment of the $\ket{n S_{1/2}}$ and $\ket{(n+1) S_{1/2}}$ differ only by a small amount
(more specifically $\bra{0} d^+ \ket{+} \approx - \bra{0} d^+ \ket{-}$), which results in $J^{+0} \approx J^{-0}$.
Thus, we can define $J^{\pm 0} = (J^{+0} + J^{-0})/2$ and ignore for now the effect of a non-vanishing $\delta^{\pm 0} = (J^{+0} - J^{-0})/2$.

Furthermore, we find that the ratio $J^{00}/J^{\pm 0}$ can be easily tuned by changing the angle $\theta$ of the applied magnetic field, as shown in \cref{fig:pairinteraction}~c).
While this in principle gives rise to an additional degree of freedom in our model, in this work we only focus on the case $J^{00} = J^{\pm 0}$.
The angle for this can be estimated by setting $3 \cos^2 \theta - 1 = 3 \sin^2 \theta$.
Taking the full effective Rydberg pair Hamiltonian into account (see \cref{app:full_hamiltonian}) we find $J^{00} = J^{\pm 0}$ to be fulfilled for $\theta \approx \qty{35.1}{^\circ}$.

As detailed in \cref{app:rydberg-to-spin}, the combined action of the three terms $J^{+0} = J^{-0} = J^{00}$ can be mapped to the spin-1 XY interaction $J^{xy}$ in \cref{eq:many-body-H}.

\subsection*{Van der Waals interactions}
In real Rydberg systems we have more than three states, which also give rise to van der Waals interactions due to the coupling to additional Rydberg states.
Often the van der Waals contributions are small compared to the dipole-dipole interactions, because the additional coupled states are energetically far detuned from the states of interest.
However, because the three states defined in \cref{eq:states} are close to a Förster resonance, an additional close by state $\ket{v} = \ket{n P_{3/2}, m_j = +1/2}$ couples to our three states of interest and thus gives rise to a significant van der Waals term.
While this is not the only van der Waals term and more states are considered in the calculation, the van der Waals term due to the state $\ket{v}$ is the dominant one and captures the physics of the additional term $V$ in \cref{eq:many-body-H}.

As shown in \cref{fig:pairinteraction}~b), the state $\ket{v}$ is detuned from the $\ket{0}$ and $\ket{\pm}$ state by an energy of $\Delta/h \approx \qty{100}{MHz}$ (at $B = \qty{60.7}{G}$).
Since the relevant couplings are on the order of $\qty{10}{MHz}$ this allows us to treat the effect of the $\ket{v}$ state perturbatively.

In second order perturbation theory, this state allows van der Waals processes like
\begin{align}
  V^\text{diag}: \quad \ket{+-} &\xrightarrow{d_i^0d_j^+} \ket{v0} \xrightarrow{d_i^0d_j^-} \ket{+-}
  \,, \nonumber \\
  V^\text{offd}: \quad \ket{+-} &\xrightarrow{d_i^0d_j^+} \ket{v0} \xrightarrow{d_i^0d_j^-} \ket{-+}
  \,,
\end{align}
where the first process gives rise to a diagonal energy term $V^\text{diag}$ and the second process to an off-diagonal interaction term $V^\text{offd}$.
Analogously, we find processes for the $\ket{-+}$ state.
Combining all terms, we find the following coupling strength
\begin{align}
  V = \frac{9 \sin^2\theta \cos^2\theta}{r^6 \Delta} \abs{\bra{v} d^0 \ket{+}}^2 \abs{\bra{0} d^+ \ket{-}}^2
  \,,
  \label{eq:V}
\end{align}
which is the same for the diagonal and off-diagonal terms $V^\text{diag} = V^\text{offd} \equiv V$.

Importantly, the van der Waals strength can be tuned with respect to the dipole-dipole interactions by changing the distance $r$ using the $1/r^6$ dependence of the van der Waals term compared to the $1/r^3$ dependence of the dipole-dipole term.
This can be seen in \cref{fig:pairinteraction}~d), where we plotted the interaction strengths $J^{\pm 0}$, $J^{00}$ and $V$ as a function of the distance $r$ between the Rydberg atoms.
Additionally, in \cref{fig:pairinteraction}~c), the angle dependence of the van der Waals term is plotted.
(Note that the van der Waals strength can in principle also be tuned by changing the detuning $\Delta$ of the state $\ket{v}$.
This can be done by applying some electric field $\vec{E}$ and changing the magnetic field $B$ accordingly, to keep the anisotropy energy $D$ fixed.)

We have calculated the interaction strengths of two Rydberg atoms by simulating the Hamiltonian (including up to $\sim 5000$ energetically nearby Rydberg pair states) with the \emph{pairinteraction} software~\cite{weberTutorialCalculationRydberg2017} (see \cref{app:full_hamiltonian} for more details).
This gives rise to additional van der Waals interactions, which also lead to a difference between $V^\text{diag}$ and $V^\text{offd}$.
In \cref{fig:pairinteraction}~c) and d), we have plotted $V = (V^\text{diag} + V^\text{offd})/2$ and the shaded area around the $V$ line indicates the difference of the diagonal and off-diagonal terms.

Notice again that, for dipole-dipole interacting systems, the van der Waals terms are typically a small perturbation.
Here, however, the contribution of the strongest van der Waals term is on the order of the dipole-dipole coupling and enriches the phase diagram of our spin-1 model.

\section{Phase diagram of the spin-1 model\label{sec:phase-diagram}}
\begin{figure}
  \centering
  \includegraphics[width=\linewidth]{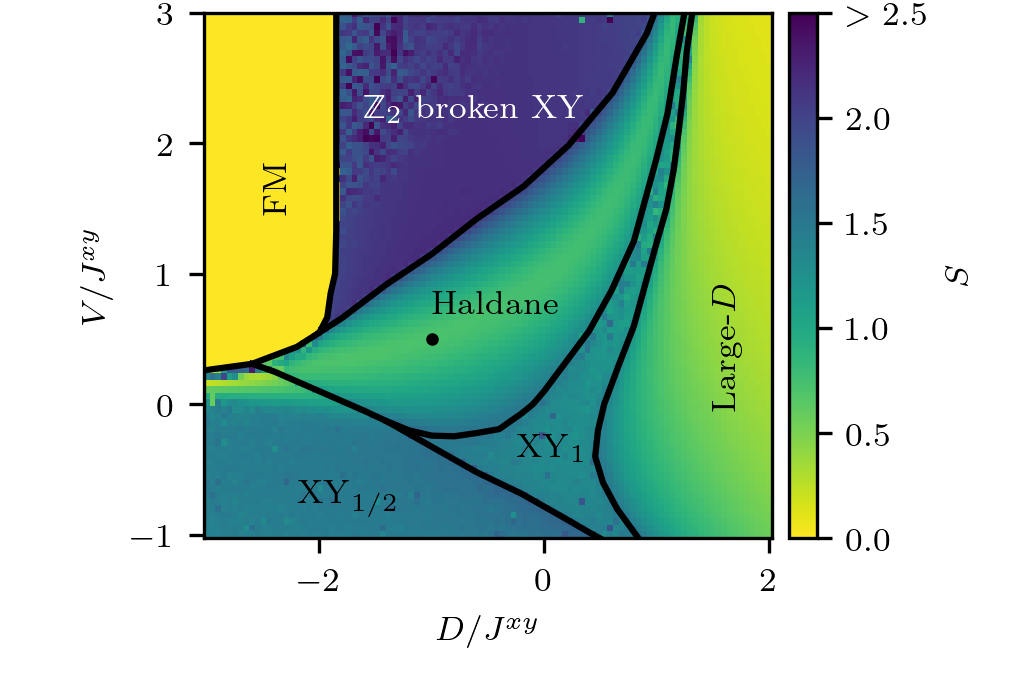}
  \caption{
    Phase diagram of our model [see \cref{eq:many-body-H}], calculated by using infinite DMRG within the \emph{TeNPy} library~\protect{\cite{hauschildEfficientNumericalSimulations2018}} and a bond dimension of $\chi = 400$.
    The color indicates the entanglement entropy $S$ of the ground state and is meant as a qualitative measure of the different phases, where local maxima indicate phase transitions.
    The solid lines are guide for the eyes and do not represent the exact phase transitions.
    For trivial product states the entropy is zero, while for gapless phases the entropy would diverge for increasing bond dimensions.
    Inside the Haldane phase one can see the entropy converging to a value of about $\log 2 \approx 0.69$, corresponding to cutting one singlet bond of the AKLT state.
  }
  \label{fig:phase-diagram}
\end{figure}
We choose $J^{xy}$ as energy scale.
Thus, we are left with two degrees of freedom and can plot the phase diagram of the Hamiltonian \eqref{eq:many-body-H} over $D$ and $V$.
By plotting the entanglement entropy $S$, as shown in \cref{fig:phase-diagram}, we can identify all occurring phase transitions (the solid lines are guide for the eyes and not exact phase transitions).
To specify and characterize the different phases, we also examined other properties, such as string order parameters and the projective representation of the symmetries, which are discussed in detail in \cref{app:phase-diagram}.
The phase diagram was calculated using infinite DMRG within the \emph{TeNPy} library~\cite{hauschildEfficientNumericalSimulations2018}.
Before we focus on the Haldane phase, let us briefly discuss the other phases in the phase diagram.

In the case of large positive values of $D$, we are in a gapped phase with a unique ground state (\emph{Large-$D$ phase}).
For $D \to \infty$, the ground state is a simple product state of all spins in the $\ket{0}$ state, thus the phase is a topological trivial phase.
The ground state is disordered and invariant under the $D_2$ and time reversal symmetry.

Next we consider the case of large negative values of $D \to - \infty$.
In this case, the $\ket{0}$ state has a large energy penalty and effectively gets frozen out, such that the Hamiltonian reduces to a two-level Hamiltonian
\begin{equation}
  H_\text{effective}
  = \sum_{i>j} \frac{V}{r_{ij}^6} \left( \sigma^+_i \sigma^-_j + \sigma^-_i \sigma^+_j \right)
  -\frac{V}{2 r_{ij}^6} \sigma^z_i \sigma^z_j\
  \,,
  \label{eq:H_S_12}
\end{equation}
with $\sigma^\pm = \ketbra{\pm}{\mp}$ and $\sigma^z = \ketbra{+}{+} - \ketbra{-}{-}$.
This is the well known spin-$1/2$ XXZ model, but with finite range interactions $\propto r_{ij}^{-6}$.
For negative values of $V$ we are in the gapless \emph{spin-1$/$2 ferromagnetic XY phase} (XY$_{1/2}$)~\cite{peterAnomalousBehaviorSpin2012}, where the low energy excitations are given by the Goldstone modes.
The ground state exhibits quasi-long range order, i.e. the off-diagonal order parameter
$\langle \sigma^+_i \sigma^-_j \rangle = \frac{1}{4} \langle (S^+_i)^2 (S^-_j)^2 \rangle$
decays algebraically with the distance $r_{ij}$.
In the same limit of large negative $D$ but for positive values of $V$, we are in the gapped \emph{Ising ferromagnetic phase} (FM).
In the whole phase the two degenerate ground states are simple product states of all spins in the $\ket{+}$ or $\ket{-}$ state.
It is a symmetry-broken phase and can easily be characterized by the Ising ferromagnetic long range order $\langle S^z_i S^z_j \rangle$.

Now, we examine the \emph{spin-1 antiferromagnetic XY phase} (XY$_{1}$), which, similarly to the XY$_{1/2}$ phase, is a gapless phase.
The ground state also exhibits quasi-long range order, but now for both off-diagonal order parameters $\langle (-1)^{i-j} S^+_i S^-_j \rangle$ and $\langle (S^+_i)^2 (S^-_j)^2 \rangle$, which decay algebraically with the distance $r_{ij}$.
Because the phase transition from the XY$_{1}$ phase to the Haldane phase is a Berezinskii-Kosterlitz-Thouless (BKT) transition~\cite{schulzPhaseDiagramsCorrelation1986,kitazawaPhaseDiagram11996,chenGroundstatePhaseDiagram2003}, it is hard to determine the exact boundaries from the correlation functions~\cite{hengsuNonlocalCorrelationsHaldane2012}.
To get the precise BKT boundaries one can either use bond dimension scaling and very large distances utilizing the infinite DMRG algorithm~\cite{hengsuNonlocalCorrelationsHaldane2012,gongKaleidoscopeQuantumPhases2016}.
Alternatively, one can do finite size scaling as described in~\cite{kitazawaCriticalProperties11997,chenGroundstatePhaseDiagram2003}.

For large positive values of $V$ we are again in a gapless phase.
The ground state breaks the $\mathbb{Z}_2$ symmetry ($\pi$ rotation around the $x$-axis) of the Hamiltonian, which can be seen in DMRG by adding a small effective magnetic field $\sum_i S^z_i$.
Furthermore, the ground state exhibits quasi-long range order in the off-diagonal order parameter $\langle (-1)^{i-j} S^+_i S^-_j \rangle$.
Thus, we call this phase a \emph{$\mathbb{Z}_2$ broken antiferromagnetic XY phase}.

We now focus on the \emph{Haldane phase} in the middle of the phase diagram.
It is a gapped phase with no long range order but finite string order parameters.
The string order parameter is defined as
\begin{equation}
  \mathcal{S}_{i,j}^\alpha =  - \bra{\psi} S^\alpha_i \left( \prod_{k=i+1}^{j-1} (-1)^{S^\alpha_k} \right) S^\alpha_j \ket{\psi}
  \,,
  \label{eq:string_order}
\end{equation}
and is a measure for the hidden antiferromagnetic order in the Haldane phase.
At the marked point, all three string order parameters $\mathcal{S}_{0,500}^{x,y,z}$ are larger than $0.4$ (note that for the perfect AKLT state $\mathcal{S}_{1,\infty}^{x,y,z} = 4/9$).
In addition to the string order parameters, we characterized the Haldane phase using the degenerate entanglement spectrum~\cite{pollmannEntanglementSpectrumTopological2010} and the invariance of the infinite MPS under the $D_2$ and time reversal symmetries, which exhibit non-trivial phase factors in their projective representations~\cite{pollmannDetectionSymmetryProtected2012}.
We also verified a significant overlap with the perfect AKLT state.
For more details on the phase diagram and the properties of the Haldane phase see \cref{app:phase-diagram}.

From now on, we focus on the point $D = - J^{xy}$ and $V = 0.6 J^{xy}$, which is marked in \cref{fig:phase-diagram} by a black dot and lies inside the Haldane phase.

\section{Experimental proposal\label{sec:exp-proposal}}
In this section, we propose a concrete example of experimental parameters that result in the desired values for $D$ and $V$ in a Rubidium setup.
We discuss the Haldane phase in more detail within this experimental setup, including finite size results, possible detection schemes, and a potential ground state preparation scheme.

Note that neither the choice of Rubidium nor the choice of $n \approx 80$ is crucial for our proposal.
Very similar Förster resonances can also be found in other Alkali atoms.
Especially in Caesium the experimental parameters to realize the Haldane phase are very similar to Rubidium.
Choosing $n \approx 80$ is a trade-off between two experimental constraints:
For smaller $n$ one needs a larger magnetic field as well as smaller interatomic distances to keep the ratio of the energies and interaction strengths the same.
On the other hand, for larger $n$ the interatomic distance has to be larger, resulting in an overall smaller energy scale $J^{xy}$.

In \cref{sec:model}, we discussed that the parameters $D$ and $V$ can be tuned in a Rydberg system by the experimental parameters.
We propose to use the following states
\begin{align}
  \ket{+} &= \ket{^{87}\text{Rb}; 81 S_{1/2}, m_j = +1/2}
  \,, \nonumber \\
  \ket{0} &= \ket{^{87}\text{Rb}; 80 P_{3/2}, m_j = +3/2}
  \,, \nonumber \\
  \ket{-} &= \ket{^{87}\text{Rb}; 80 S_{1/2}, m_j = +1/2}
  \,.
  \label{eq:exp_prop_states}
\end{align}
The atoms are placed with an interatomic distance of $r = \qty{11.6}{\mu m}$
and a magnetic field of strength $B = \qty{60.7}{G}$, which is aligned at an angle of $\theta = \qty{35.1}{^\circ}$ with respect to the interatomic axis.
The electric field is kept at $0$.
Those parameters result in an anisotropy energy of $D/h = \qty{-4.47}{MHz}$ and nearest-neighbor ($nn$) interaction strengths of $J^{xy}_{nn}/h = \qty{4.47}{MHz}$ and $V_{nn}/h = \qty{2.68}{MHz}$.
This corresponds to the point $D = - J^{xy}_{nn}$ and $V_{nn} = 0.6 J^{xy}_{nn}$, which we motivated in the last section.
By also looking at further neighbor values, see \cref{app:full_hamiltonian} for a full list of all interaction strengths, we find that the interaction strengths decay with $1 / r_{ij}^{\alpha}$, where $\alpha \approx 3$ for dipolar interactions and $\alpha \approx 6$ for the van der Waals interactions, as predicted in \cref{sec:model}.

Note two differences with respect to the simplified model we introduced in \cref{sec:model} and discussed in \cref{sec:phase-diagram}.
First, the couplings $J^{+0}_{nn}/h = \qty{4.36}{MHz}$ and $J^{-0}_{nn}/h = \qty{4.59}{MHz}$ are not exactly the same anymore.
Second, there are now other additional van der Waals terms, see \cref{app:full_hamiltonian} for more details.
Both these effects result in a small symmetry breaking of the $D_2$ and time reversal symmetry, which however is only of the order of $0.03 J^{xy}_{nn}$ and thus results in a small splitting of the four ground states of the Haldane phase.
However, as long as these symmetry breaking terms are weak compared to the gap the features of the Haldane phase are still visible.
In the following, we always consider the full Hamiltonian including all terms listed in \cref{app:full_hamiltonian}.

Let us now investigate an experimentally realizable finite size system with these parameters and discuss possibilities to measure the Haldane phase and its properties in an experiment.
All results in this section are calculated via DMRG and tensor network time-evolution methods within the \emph{TeNPy} library~\cite{hauschildEfficientNumericalSimulations2018, zaletelTimeevolvingMatrixProduct2015}.
We use a system size of $L=10$ atoms, which already is enough to clearly see the separated edge states of the Haldane phase.
However, the results in this section also generalize to larger system sizes.
As expected, we do find four ground states with total magnetizations $M_\mathrm{tot} = -1, 0, 0, +1$ and a gap above the ground state manifold to the excited states of about $\sim \qty{600}{kHz}$.
The ground state manifold is split by a small amount of $\sim \pm \qty{70}{kHz}$, which can be explained by the symmetry breaking terms in the Hamiltonian.
\begin{figure}
  \centering
  \includegraphics{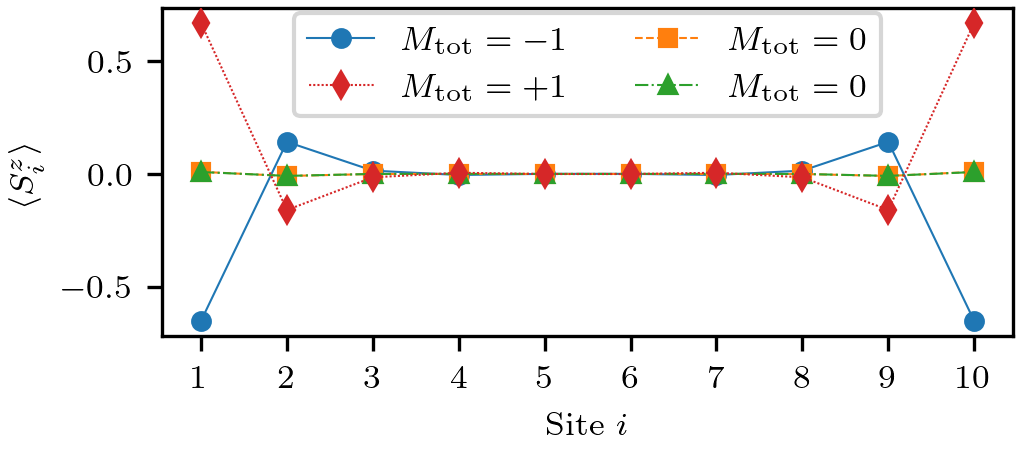}
  \caption{
  The expectation value of the local magnetization at each site for the four different ground states.
  For the $M_\mathrm{tot} = \pm 1$ states we can clearly see that the magnetization is localized at the edges.
  For the two degenerate $M_\mathrm{tot} = 0$ states the edge states hybridize due to the small system size.
  Nevertheless, for all four ground states the edge spin correlation is non-zero $\abs{\expvalue{S^z_1 S^z_{10}}} \approx 0.4$, while it vanishes in the bulk $\abs{\expvalue{S^z_1 S^z_{6}}} \approx 0$.
  This reveals the edge states for all four ground states.
  }
  \label{fig:magnetization}
\end{figure}
Looking at the site-resolved magnetization of the four different ground states in \cref{fig:magnetization}, we see the fractional edge magnetization.
Note that, even for this small system size, the edge states are already well separated from each other and, when summing over the magnetization of the 3 sites closest to the edge, we get values of almost exactly $\pm 0.5$.

\begin{figure}
  \centering
  \includegraphics{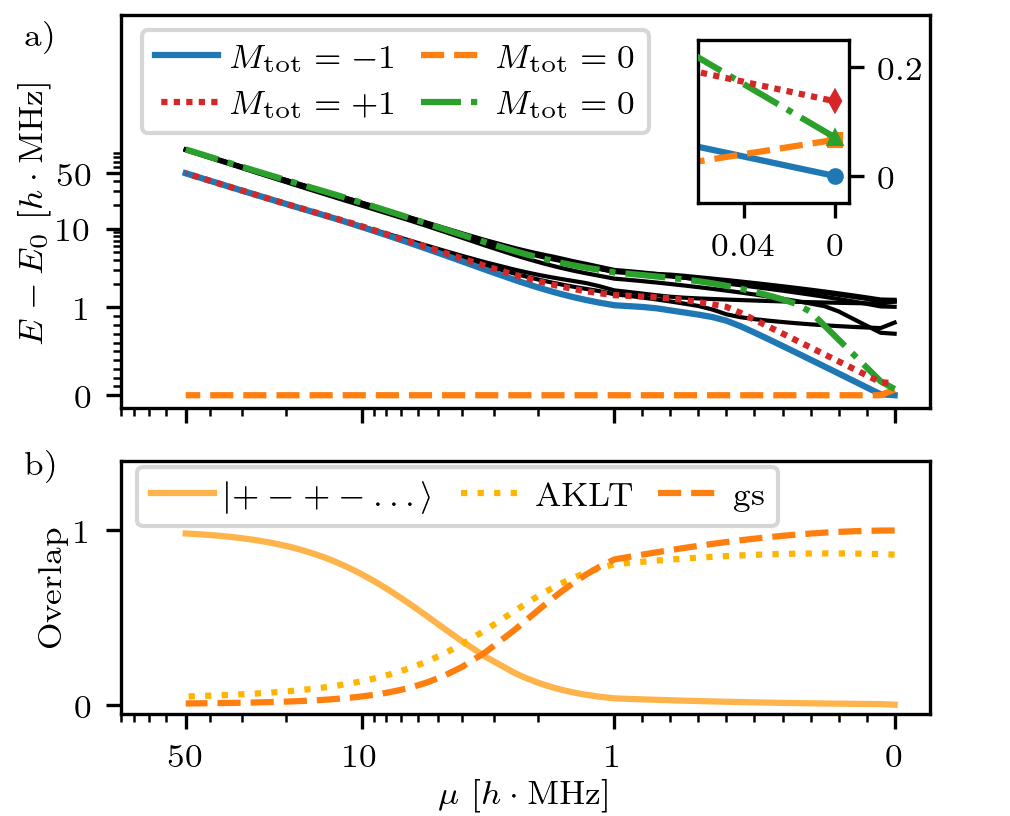}
  \caption{
    Adiabatic preparation scheme on a symlog scale (i.e. linear from $0$ to $1$ and logarithmic for the rest) for a Rydberg chain of $L = 10$ atoms.
    a) The energy spectrum ${E - E_0}$ of the ten lowest eigenstates compared to the ground state energy $E_0$ is shown for the sweep process.
    The sweep starts with a unique ground state and ends up with the fourfold near-degenerate ground state manifold of the Haldane phase.
    The four colored states correspond to the states ending in the ground state manifold of the final Hamiltonian.
    All other states are well separated from the ground state during the entire sweep by a gap of at least \qty{600}{kHz}.
    The inset shows the energy splitting of the ground state manifold at the end of the sweep.
    b) The overlap $\abs{\braket{\Psi_\mathrm{GS}|\Phi}}^2$ of the instantaneous ground state during the sweep (orange line in a)) with i) the initial product state $\ket{+-+-+-\dots}$, ii) the perfect AKLT state with edge degrees of freedom $(\uparrow, \downarrow)$ and iii) the final ground state with $M_\mathrm{tot} = 0$ of the Hamiltonian at $\mu = 0$.
  }
  \label{fig:preparation}
\end{figure}
We now discuss a method to prepare these ground states of the Haldane phase by adiabatically sweeping from a product state to the desired ground state.
Since the Haldane phase is a symmetry protected topological phase, this can only be achieved by breaking the protecting symmetries (i.e. the $D_2$ and time reversal symmetry).
An intuitive approach using a homogenous effective magnetic field as well as a Rabi frequency was already experimentally demonstrated for the SSH chain~\cite{deleseleucObservationSymmetryprotectedTopological2019}.
However, as we outline in \cref{app:preparation}, this method does not work for the AKLT state and our system.
Here, we propose a different adiabatic preparation scheme only using a staggered, effective magnetic field, similar as described in~\cite{garcia-ripollImplementationSpinHamiltonians2004}.

We start with the initial state in a product state $\ket{+-+-+-\dots}$, which can be prepared by applying spatially dependent light shifts on the Rydberg chain~\cite{chenContinuousSymmetryBreaking2023}.
By adding a strong staggered, effective magnetic field $\mu \sum_{i=1}^L (-1)^{i} S^z_i$ to the Hamiltonian, we ensure that this staggered state is initially the ground state of the system.
Now, continuously turning off this staggered, effective magnetic field $\mu$, we end up with the desired Hamiltonian.
As shown in \cref{fig:preparation}~a), the ground state manifold during this sweep is always well separated from the excited states by a gap of about $\qty{600}{kHz}$.
This allows us to adiabatically sweep from the product state to the Haldane phase, while never going through a gap closing.
In \cref{fig:preparation}~b), we plotted the overlap of the instantaneous ground state with the product state $\ket{+-+-+-\dots}$, the AKLT state, and the final ground state during the preparation sweep.
As described above, the initial ground state has a large overlap with the product state, while the final state has a large overlap with the AKLT state.
Note that the ground state of the final Hamiltonian is close to but not exactly the perfect AKLT state.
We numerically checked that for a total sweep time of $\qty{5}{\mu s}$, which is much smaller than the Rydberg lifetime of $\sim \qty{100-200}{\mu s}$ for $n=80$, the fidelity of the prepared state with the desired final ground state is larger than $0.98$.
Importantly, because the energy gap does not depend on the system size, which we also confirmed numerically, this preparation scheme also works for larger systems.

The Hamiltonian during the whole sweep conserves the total magnetization $M_\mathrm{tot}$.
This allows us to also prepare the ground states with $M_\mathrm{tot} = \pm 1$ by using odd system sizes and a product state $\ket{+- \dots -+}$ or $\ket{-+ \dots +-}$.

\begin{figure}
  \centering
  \includegraphics{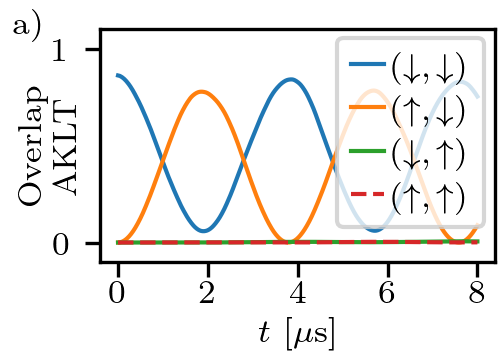}
  \includegraphics{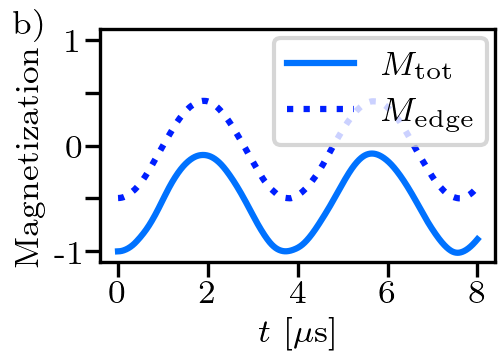}
  \caption{
    At time $t=0$ we start with the ground state of our Hamiltonian and apply a rotation around the $x$-axis on the first half of the chain (i.e. the $5$ leftmost sites) for a time $t$. The Rabi frequency is $\Omega/h = \qty{0.25}{MHz}$, which results in a $\pi$ rotation after \qty{2}{\mu s}.
    a) The overlap of the time evolved state with the four different AKLT states, which are labelled according to their edge degrees of freedom.
    b) The magnetization of the full system $M_\mathrm{tot} = \expvalue{\sum_{i=1}^{10} S^z_i}$ and of the 3 leftmost edge sites
    $M_\mathrm{edge} = \expvalue{\sum_{i=1}^{3} S^z_i}$.
  }
  \label{fig:rotation}
\end{figure}
After preparing the ground state of the Haldane phase, one can now measure the ground state properties of the Haldane phase.
These include the site-resolved magnetization shown in \cref{fig:magnetization}, where the edge states can be seen.
Furthermore, one can also measure the string order parameters, which were defined in \cref{eq:string_order}.
For this finite size system the string order parameters of the ground state are $\mathcal{S}_{3, 8}^z = 0.29$ and $\mathcal{S}_{3, 8}^x = 0.49$, showing the strong hidden antiferromagnetic order of the system.

While some features of the edge states can be seen in the static expectation values of the ground state magnetization, we also investigate the dynamics of these edge states.
This allows us to see the fractionalization of the spin-1 chain to spin-$1/2$ degrees of freedom at the edges.
We start in the ground state with total magnetization $M_\mathrm{tot} = -1$.
Then, we coherently drive the system between two different ground states by applying a microwave pulse with frequency $\omega = (E_+ - E_-) / 2$ only to the left half of the chain.
This drives both the transitions $\ket{+} \leftrightarrow \ket{0}$ and $\ket{-} \leftrightarrow \ket{0}$.
In spin language this corresponds to the $S^x$ operator, which generates rotations around the $x$-axis.
By using a Rabi frequency $\Omega$, which is small compared to the energy gap of the system, we ensure that the system is driven adiabatically and always stays inside the ground state manifold.

We now look at the time evolved state after some time $t$.
As shown in \cref{fig:rotation}~a), the state oscillates between the two AKLT states with the spin-$1/2$ degree of freedom on the left edge pointing down or up.
These oscillations correspond to Rabi oscillations of the fractionalized spin-$1/2$ degree of freedom at the left edge.
One can also observe this in an experiment by measuring the expectation value of the magnetization at the 3 leftmost sites.
As shown in \cref{fig:rotation}~b), the edge magnetization in $z$ direction oscillates between $-0.5$ and $+0.5$.

Additionally, in \cref{fig:rotation}~b), the total magnetization is plotted, showing oscillations between $-1$ and $0$.
This indicates that after an effective $\pi$ rotation (i.e., after $\qty{2}{\mu s}$ for a microwave pulse with $\Omega = \qty{0.25}{MHz}$) of half the spin-1 chain, the total magnetization of the system changes from $-1$ to $0$.
This is a clear signature of the fractionalization of the spin-1 chain into effective spin-$1/2$ edge states.
In contrast, applying a $\pi$ rotation to a single spin-1 degree of freedom would only change the magnetization by $0$ or $\pm 2$ (since $\exp(i \pi S^x) \ket{0} = - \ket{0}$ and $\exp(i \pi S^x) \ket{\pm} = \ket{\mp}$).
Importantly, when driving the system adiabatically (i.e., with a Rabi frequency $\Omega$ small compared to the gap), the state after the $\pi$ rotation remains an eigenstate of the total magnetization.

\section{Conclusion and Outlook\label{sec:conclusion}}
In this work we proposed a new way of using Förster resonances in Rydberg atoms to implement effective spin-1 systems with highly tunable interactions.
The phase diagram of this Hamiltonian exhibits a large and stable region with the Haldane phase, which is characterized by a hidden antiferromagnetic order and fractionalized edge states.
The aforementioned Haldane phase can be realized in Rydberg systems with an experimental setup and magnetic field strengths that are well within the possibilities of present-day experiments.
In addition, we discussed an adiabatic available preparation scheme utilizing experimental techniques to prepare the Haldane phase.
Finally, we studied a finite size realization of such Rydberg chains and proposed multiple ways of measuring the edge state features of the Haldane phase, including the fractionalization of these edge states.
Therefore, we have laid out a complete path for implementing, preparing, and measuring the spin-1 Haldane phase in Rydberg platforms.
Our detailed proposal should pave the way for experiments to study the spin-1 Haldane phase, and thus allows to detect and study the fractionalization of edge spin degrees of freedom.

\begin{acknowledgments}
  We would like to thank Chew Torii and Jakob Hartmann for insightful discussions.
  We acknowledge financial support by the Baden-Württemberg Stiftung via BWST Grant No. ISF2019-017 under the program Internationale Spitzenforschung.
  A.B. has received funding under Horizon Europe programme HORIZON-CL4-2022-QUANTUM-02-SGA via the project 101113690 (PASQuanS2.1), the European Research Council (ERC) Advanced Grant No. 101018511 (ATARAXIA).
  J.P. acknowledges support from the QuantERA II Programme (Mf-QDS) that has received funding from the European Union's Horizon 2020 research and innovation programme under Grant Agreement No 101017733, and from grant TED2021-130578B-I00 and grant PID2021-124965NB-C21 funded by MICIU/AEI/10.13039/501100011033 and by the European Union NextGenerationEU/PRTR.
\end{acknowledgments}

\section*{Data Availability}
The data that support the findings of this article are openly available~\cite{DARUS-4990_2025}.
\nocite{DARUS-4990_2025}

\appendix
\section{Mapping of Rydberg interactions to Spin-1 operators\label{app:rydberg-to-spin}}
In \cref{sec:model}, we discussed the Rydberg energies and interactions in our model, which are again summarized here
\begin{alignat}{2}
  &H^{D}_i
  &&= D \left( \ketbra{+}{+} + \ketbra{-}{-} \right)
  \,, \nonumber \\
  &H^\mathrm{XY}_{ij}
  &&= \Bigl[ J^{+0} \ketbra{0+}{+0} + J^{-0} \ketbra{-0}{0-}
    \nonumber \\
    & && \quad \quad
    + J^{00} \left( \ketbra{-+}{00} + \ketbra{+-}{00} \right) \Bigr] + \hc
  \,, \nonumber \\
  &H^{\mathrm{V}}_{ij}
  &&= V^\text{offd} \left( \ketbra{+-}{-+} + \ketbra{-+}{+-} \right)
  \nonumber \\
  & && \quad \quad
  + V^\text{diag} \left( \ketbra{+-}{+-} + \ketbra{-+}{-+} \right)
  \,.
  \label{eq:H:parts}
\end{alignat}
Note that the anisotropy term $H^{D}_i$ is due to the Förster resonance offset of the state $\ket{00}$ relative to the state $\ket{+-}$ and, up to a global energy shift, we could also write it in the more intuitive form $H^{D}_i = - D \ketbra{0}{0}$.

Here, we map these Rydberg interactions to spin-1 interactions and the spin-1 algebra.
The spin-1 operators in the eigenbasis of $S^z$, i.e. $\ket{+}$, $\ket{0}$ and $\ket{-}$, are given as
\begin{align}
  S^x &= \frac{1}{\sqrt{2}} \begin{pmatrix} 0 & 1 & 0 \\ 1 & 0 & 1 \\ 0 & 1 & 0 \end{pmatrix}
  \, ,
  S^y = \frac{1}{\sqrt{2}} \begin{pmatrix} 0 & -i & 0 \\ i & 0 & -i \\ 0 & i & 0 \end{pmatrix}
  \, , \nonumber \\
  S^z &= \begin{pmatrix} 1 & 0 & 0 \\ 0 & 0 & 0 \\ 0 & 0 & -1 \end{pmatrix}
  \, \text{ and }
  S^\pm = S^x \pm i S^y
  \,.
\end{align}
One can easily see that the anisotropy term simply maps to $H^{D}_i = D (S^z_i)^2$.

As discussed in \cref{sec:model}, the dipole-dipole interactions can be tuned to be approximately the same: $J^{+0} = J^{-0} = J^{00} \equiv J^{xy}$.
This simplifies the expression, since $H^\mathrm{XY}_{ij}$ now simply involves all total spin conserving interactions, which change the spin of each individual site by $\pm 1$.
One can quickly convince oneself that this corresponds to the following spin-1 XY interactions
\begin{equation}
  H^\mathrm{XY}_{ij} = \frac{J^{xy}}{2} \left( S^+_i S^-_j + S^-_i S^+_j \right) = J^{xy} \left( S^x_i S^x_j + S^y_i S^y_j \right)
  \,.
\end{equation}
For the van der Waals term, we first look at the off-diagonal elements $\ketbra{+-}{-+} + \ketbra{-+}{+-}$.
These are total spin conserving interactions, which change the spin magnetization of each individual site by $\pm 2$.
We therefore can describe this term by $(S^+_i)^2 (S^-_j)^2 + (S^-_i)^2 (S^+_j)^2$.
The diagonal elements only act on states if neither of them is $\ket{0}$ leading to $S^z_i S^z_j$, and they have not the same magnetization, which can be enforced by $(S^z_i S^z_j - 1)$.
Including the correct prefactors we thus end up with
\begin{align}
  H^{\mathrm{V}}_{ij}
  &= \frac{V^\text{offd}}{4} \left( (S^+_i)^2 (S^-_j)^2 + (S^-_i)^2 (S^+_j)^2 \right)
  \nonumber \\
  & \quad
  + \frac{V^\text{diag}}{2} S^z_i S^z_j (S^z_i S^z_j - 1)
  \,.
\end{align}

\section{More details on the phase diagram\label{app:phase-diagram}}
\begin{figure}
  \centering
  \includegraphics{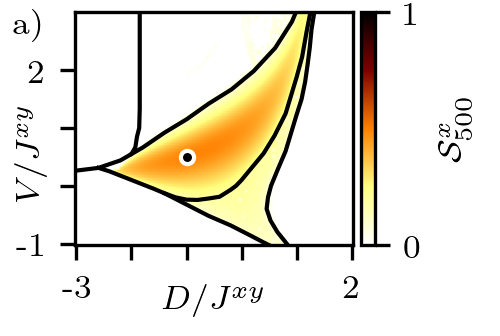}
  \includegraphics{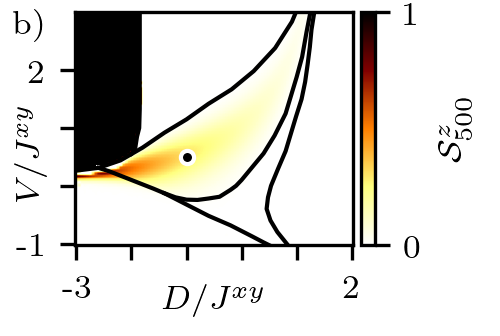}
  \includegraphics{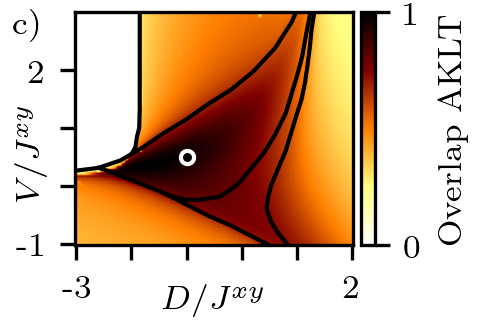}
  \includegraphics{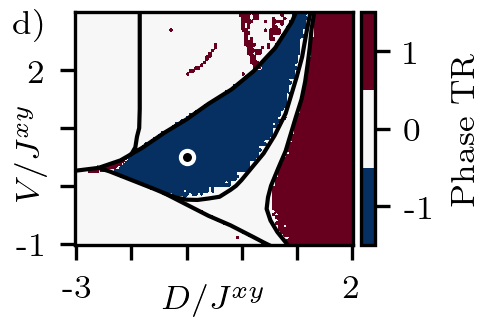}
  \caption{
    The same phase diagram as discussed in \cref{sec:phase-diagram} but now showing
    a) and b) the $x$ and $z$ string order parameter
    c) the overlap of the ground state with the perfect AKLT state
    d) the phase factor of the projective representation of the time reversal symmetry.
    All plots have a resolution of $100 \times 100$ points.
  }
  \label{fig:phase-diagram-details}
\end{figure}
In this section we elaborate on the phase diagram discussed in \cref{sec:phase-diagram}, see also~\cite{MScJohannesMoegerle}.
Specifically, we discuss the properties of the Haldane phase that can be used to numerically distinguish it from the other phases.

In \cref{fig:phase-diagram-details}~a)~and~b), we plotted the string order parameters $\mathcal{S}_{0,500}^x$ and $\mathcal{S}_{0,500}^z$ for two spins separated by $500$ sites.
The $x$ string order parameter has a finite value in the whole Haldane phase as well as a smaller but finite value also in the XY$_{1}$ phase, which is due to the finite separation and finite bond dimensions and the small decay of the $x$ string order parameter in the XY$_{1}$ phase (for more details about this see \cite{hengsuNonlocalCorrelationsHaldane2012}).
The $z$ string order parameter is perfectly $1$ in the ferromagnetic phase and, apart from that, only has finite values inside the Haldane phase.
However, it gets smaller the closer we get to any phase transition.

Next, in \cref{fig:phase-diagram-details}~c), we have plotted the largest eigenvalue of the transfer matrix of the contraction of the perfect AKLT unit cell and the unit cell of the infinite MPS ground state, where the unit cells consisted of two single sites.
This effectively describes an overlap of the ground state with the AKLT state per unit cell (for an infinite system).
One can see that this overlap peeks inside the Haldane phase and reaches values close to $1$, thus letting us conclude that the ground state is indeed very close to the AKLT state.

Finally, in \cref{fig:phase-diagram-details}~d), we plotted the phase factor of the projective representation of the time reversal (TR) symmetry.
This is defined to be $0$ where the ground state is not invariant under time reversal and otherwise reflects the phase factor of the projective representation of the time reversal symmetry group on the infinite MPS if the state is invariant (for more details on this see \cite{pollmannDetectionSymmetryProtected2012}).
We can see that the Large-$D$ phase is a symmetric, but topologically trivial phase (i.e. phase factor of $+1$) under time reversal.
On the other hand, the ground state in the Haldane phase is also invariant under time reversal, but has a non-trivial phase factor $-1$ and thus corresponds to the topological counterpart of the symmetric Large-$D$ phase.

\section{Full experimental Rydberg Hamiltonian\label{app:full_hamiltonian}}
For the full experimental Hamiltonian, we consider pairs of Rydberg atoms with distances ranging from nearest neighbors ($r = \qty{11.6}{\mu m}$) up to fifth-nearest neighbors ($r = \qty{58.0}{\mu m}$).
For each pair, we use the \emph{pairinteraction} software~\cite{weberTutorialCalculationRydberg2017} to calculate an effective interaction Hamiltonian for the nine pair states of interest ($\ket{++}$, $\ket{+0}$, $\dots$, $\ket{--}$).
This is done by first considering the single Rydberg atoms and taking into account all states close to the $80S$, $80P$ and $81S$ manifolds (i.e. states within an energy window of $\pm \qty{80}{GHz}$, a principal quantum number $76 \leq n \leq 85$ and an azimuthal quantum number $l \leq 4$).
We then use the eigenstates of the single atom Hamiltonian (including the magnetic field) to construct a pair Hamiltonian.
However, we restrict the basis of the pair Hamiltonian to only pair states within an energy window of $\pm \qty{2}{GHz}$ around the five different subspaces of pair magnetization $M_\text{pair} = \pm 2, \pm 1, 0$, since the effect of states further away is negligible.
This already results in up to $\sim 5000$ pair states that we need to consider for each subspace.
Finally, we apply a direct rotation that maps the pair Hamiltonian to the subspace of interest and thus obtain an effective interaction Hamiltonian~\cite{bravyiSchriefferWolffTransformationQuantum2011}.
Combining the 5 subspaces to a full Hamiltonian, we then obtain a $9 \times 9$ interaction matrix for each pair of the following structure
  {\small
    \begin{equation}
      H_{ij} =
      \begin{pmatrix}
        V^{++}_{ij} & \mzero      & \mzero             & \mzero      & \mzero      & \mzero      & \mzero             & \mzero      & \mzero      \\
        \mzero      & V^{+0}_{ij} & \mzero             & J^{+0}_{ij} & \mzero      & \mzero      & \mzero             & \mzero      & \mzero      \\
        \mzero      & \mzero      & V^\text{diag}_{ij} & \mzero      & J^{00}_{ij} & \mzero      & V^\text{offd}_{ij} & \mzero      & \mzero      \\
        \mzero      & J^{+0}_{ij} & \mzero             & V^{+0}_{ij} & \mzero      & \mzero      & \mzero             & \mzero      & \mzero      \\
        \mzero      & \mzero      & J^{00}_{ij}        & \mzero      & V^{00}_{ij} & \mzero      & J^{00}_{ij}        & \mzero      & \mzero      \\
        \mzero      & \mzero      & \mzero             & \mzero      & \mzero      & V^{-0}_{ij} & \mzero             & J^{-0}_{ij} & \mzero      \\
        \mzero      & \mzero      & V^\text{offd}_{ij} & \mzero      & J^{00}_{ij} & \mzero      & V^\text{diag}_{ij} & \mzero      & \mzero      \\
        \mzero      & \mzero      & \mzero             & \mzero      & \mzero      & J^{-0}_{ij} & \mzero             & V^{-0}_{ij} & \mzero      \\
        \mzero      & \mzero      & \mzero             & \mzero      & \mzero      & \mzero      & \mzero             & \mzero      & V^{--}_{ij}
      \end{pmatrix}
      \,.
    \end{equation}
  }
The interaction strengths for the parameters defined in \cref{sec:exp-proposal} for each pair up to a distance of 5 sites are listed in the following \cref{tab:H:params}.
Note that values smaller than $\qty{0.01}{MHz}$ are not listed (for $V^{00}$ this is true even for nearest neighbors).
\begin{table}[H]
  \centering
  \caption{Interaction strengths for the Rydberg Hamiltonian~\eqref{eq:exp:H} in $\unit{h \cdot \mathrm{MHz}}$.}
  \begin{tabular}{c|c|c|c|c|c}
    $\abs{i-j}$            & 1    & 2    & 3    & 4    & 5    \\
    \hline \hline
    $J^{+0}_{ij}/h$        & 4.36 & 0.56 & 0.17 & 0.07 & 0.04 \\
    $J^{-0}_{ij}/h$        & 4.59 & 0.59 & 0.17 & 0.07 & 0.04 \\
    $J^{00}_{ij}/h$        & 4.46 & 0.57 & 0.17 & 0.07 & 0.04 \\
    $V^\text{diag}_{ij}/h$ & 2.99 & 0.05 &      &      &      \\
    $V^\text{offd}_{ij}/h$ & 2.38 & 0.04 &      &      &      \\
    $V^{+0}_{ij}/h$        & 1.10 & 0.02 &      &      &      \\
    $V^{-0}_{ij}/h$        & 1.10 & 0.02 &      &      &      \\
    $V^{++}_{ij}/h$        & 1.84 & 0.03 &      &      &      \\
    $V^{--}_{ij}/h$        & 2.13 & 0.03 &      &      &      \\
  \end{tabular}
  \label{tab:H:params}
\end{table}
The anisotropy energy $D$ is simply given by half the energy difference between the $\ket{+-}$ and $\ket{00}$ states and for the parameters given in \cref{sec:exp-proposal} is $D/h = \qty{-4.47}{MHz}$.
The full many-body Hamiltonian can then be written as
\begin{align}
  H &= \sum_{i < j} H_{ij} + D \sum_{i} \left( S^z_i \right)^2
  \,.
  \label{eq:exp:H}
\end{align}

\section{Details on the preparation scheme\label{app:preparation}}
\begin{figure}
  \centering
  \includegraphics{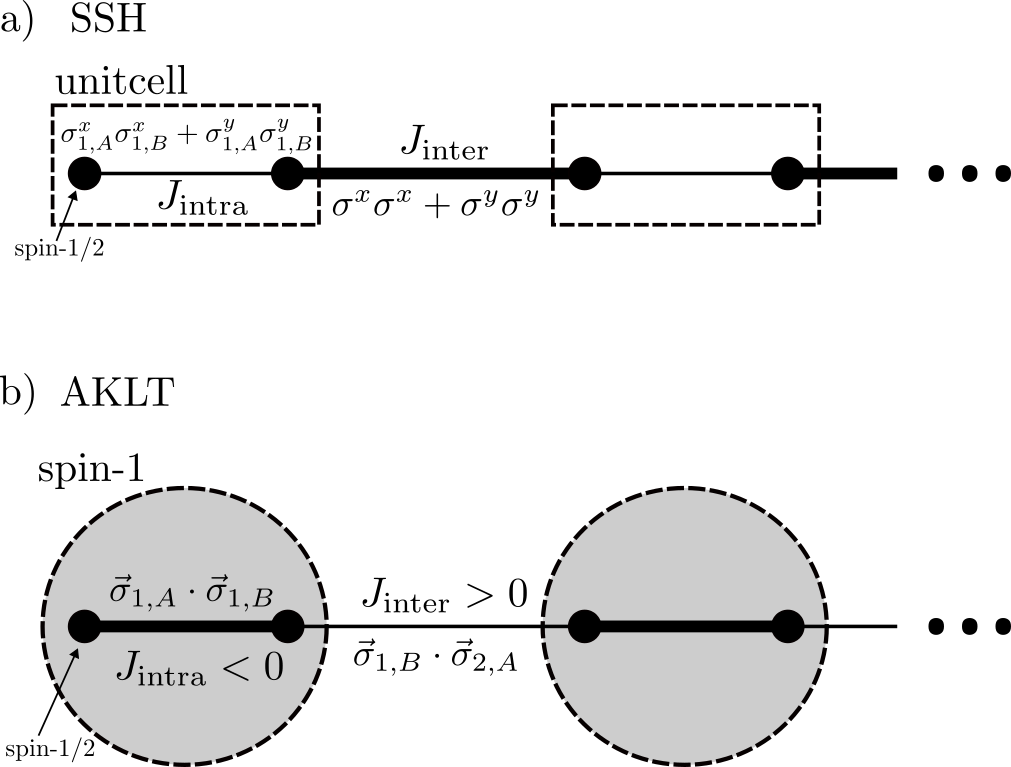}
  \caption{
    Sketch of the SSH and AKLT model with their typical interactions, where thick lines indicate strong couplings.
    a) The SSH model with strong inter-cell XY couplings and weak intra-cell XY couplings, where each unit-cell consists of two spin-$1/2$ particles.
    b) The AKLT model, where each spin-1 particle can be thought of as a pair of strongly ferromagnetic coupled virtual spin-$1/2$ particles.
    In this picture the AKLT interactions correspond to isotropic, antiferromagnetic couplings between those virtual spin-$1/2$ particles of neighboring spin-1 particles.
    \label{fig:ssh-aklt}
  }
\end{figure}
The SSH chain is in the same symmetry protected topological phase as the AKLT model.
Therefore, to adiabatically connect the SSH chain to a trivial product state, one has to break the protecting symmetries.
In this section, we review an adiabatic preparation scheme for the SSH chain, which uses a homogenous microwave field to break the protecting symmetries.
This scheme was already successfully realized in a system of Rydberg atoms~\cite{deleseleucObservationSymmetryprotectedTopological2019}.
Then, we discuss how this scheme changes when considering an antiferromagnetic SSH chain instead of the ferromagnetic one
(similar differences for the ferromagnetic and antiferromagnetic case have also been discussed for preparing 2d XY magnets~\cite{chenContinuousSymmetryBreaking2023}).
Finally, we then connect the antiferromagnetic SSH chain to the AKLT model and discuss the adapted preparation scheme in the context of the AKLT model.
By doing so, we motivate the preparation scheme proposed in \cref{sec:exp-proposal}, which only uses a staggered, effective magnetic field.

The periodic SSH Hamiltonian is defined as
\begin{align}
  H_\mathrm{SSH}
  &= \sum_{i=1}^{L} J_\mathrm{intra} \left( \sigma^x_{i,A} \sigma^x_{i,B} + \sigma^y_{i,A} \sigma^y_{i,B} + \delta \sigma^z_{i,A} \sigma^z_{i,B} \right)
  \nonumber \\
  &+ \sum_{i=1}^{L} J_\mathrm{inter} \left( \sigma^x_{i,B} \sigma^x_{i+1,A} + \sigma^y_{i,B} \sigma^y_{i+1,A} + \delta \sigma^z_{i,B} \sigma^z_{i+1,A} \right)
  \,,
  \label{eq:H-SSH}
\end{align}
where $\sigma^\alpha_{L+1,A} = \sigma^\alpha_{1,A}$.
$J_\mathrm{intra}$ describes the coupling within one unit cell between the sites $A$ and $B$, and $J_\mathrm{inter}$ describes the coupling of site $B$ and site $A$ between two neighboring unit cells.
$\delta$ is an anisotropy parameter and is only used for connecting the SSH Hamiltonian to the AKLT Hamiltonian.
For the SSH chain, the inter-cell coupling is typically stronger than the intra-cell coupling, i.e. $\abs{J_\mathrm{inter}} > \abs{J_\mathrm{intra}}$.
These interactions are also sketched in \cref{fig:ssh-aklt}~a), where thick lines indicate strong couplings.

We first consider the ferromagnetic case $J_\mathrm{inter} < 0$ with $\abs{J_\mathrm{intra}} \ll \abs{J_\mathrm{inter}}$ and $\delta = 0$ (this is similar to the SSH model used in~\cite{deleseleucObservationSymmetryprotectedTopological2019}).
This corresponds to strong ferromagnetic inter-cell couplings, which ground state can be described in the limit of $J_\mathrm{intra} \to 0$ by a product of neighboring triplet states
\begin{equation}
  \ket{\psi_\mathrm{SSH,ferro}} \approx \prod_{i=1}^{L} \frac{1}{\sqrt{2}} \left( \ket{\uparrow}_{i,B} \ket{\downarrow}_{i+1,A} + \ket{\downarrow}_{i,B} \ket{\uparrow}_{i+1,A} \right)
  \,.
\end{equation}
The crucial step to adiabatically prepare this state is to have in addition to the SSH Hamiltonian a strong, homogenous microwave field $- \Omega \sum_{i, a \in \{A,B\}} \sigma^x_{i, a}$,
for which the ground state is a simple product state $\ket{\Phi_\mathrm{homog}} = \prod_{i, a \in \{A,B\}} \ket{\rightarrow}_{i,a}$, where $\ket{\rightarrow} = \frac{1}{\sqrt{2}} \left( \ket{\uparrow} + \ket{\downarrow} \right)$.
Then, slowly turning the microwave field off yields the desired SSH ground state.
Note that the product state $\ket{\Phi_\mathrm{homog}}$ has a large overlap with the ferromagnetic SSH ground state $\ket{\psi_\mathrm{SSH,ferro}}$, which is an intuitive way to explain why during this preparation scheme no gap closing occurs.

Secondly, we consider the antiferromagnetic case $J_\mathrm{inter} > 0$, again with $\abs{J_\mathrm{intra}} \ll \abs{J_\mathrm{inter}}$ and $\delta = 0$.
The ground state for these strong antiferromagnetic inter-cell couplings in the limit of $J_\mathrm{intra} \to 0$ is given by the product state of neighboring singlet states
\begin{equation}
  \ket{\psi_\mathrm{SSH,antiferro}} \approx \prod_{i=1}^{L} \frac{1}{\sqrt{2}} \left( \ket{\uparrow}_{i,B} \ket{\downarrow}_{i+1,A} - \ket{\downarrow}_{i,B} \ket{\uparrow}_{i+1,A} \right)
  \,.
\end{equation}
Contrary to the ferromagnetic case, the overlap of this antiferromagnetic SSH ground state with the product state $\ket{\Phi_\mathrm{homog}}$ is zero.
And indeed, checking the energy spectrum of the Hamiltonian when turning off the homogenous microwave field $\Omega$, we find a gap-closing point during the sweep.

Notice however that the antiferromagnetic SSH Hamiltonian can be obtained from the ferromagnetic SSH Hamiltonian by rotating all spins of the sublattice $B$ by $\pi$ around the $z$-axis.
Thus, we can instead also switch the sign of the microwave field for the sublattice $B$ resulting in a staggered microwave field.
The initial state for this staggered microwave field is then also a product state $\ket{\Phi_\mathrm{staggered}} = \prod_{i} \ket{\rightarrow}_{i,A} \ket{\leftarrow}_{i,B}$, where $\ket{\leftarrow} = \frac{1}{\sqrt{2}} \left( \ket{\uparrow} - \ket{\downarrow} \right)$.
This state again has a large overlap with the antiferromagnetic SSH ground state $\ket{\psi_\mathrm{SSH,antiferro}}$ and allows for a gapless preparation path.

Now, we relate the antiferromagnetic SSH spin-$1/2$ chain to the spin-1 AKLT model.
Adding $\delta = 1$ to the antiferromagnetic SSH Hamiltonian and a strong ferromagnetic intra-cell coupling $- J_\mathrm{intra} \gg J_\mathrm{inter}$, we can connect the SSH Hamiltonian to the AKLT Hamiltonian along a gapped path (see appendix of~\cite{deleseleucObservationSymmetryprotectedTopological2019} for more details).
The AKLT Hamiltonian is given by
\begin{align}
  H_\mathrm{AKLT}
  &= \sum_{i=1}^{L} P^{S=2}_{i,i+1}
  \\
  &\sim \sum_{i=1}^{L} J_\mathrm{intra} \vec{\sigma}_{i,A} \vec{\sigma}_{i,B} + J_\mathrm{inter} \vec{\sigma}_{i,B} \vec{\sigma}_{i+1,A}
  \,,
\end{align}
where $P^{S=2}_{i,i+1}$ is the projector of two neighboring spin-1 particles onto a spin-2 subspace.
Here, the strong ferromagnetic intra-cell coupling $J_\mathrm{intra} < 0$ of two spin-$1/2$ particles inside one unit cell projects each unit cell effectively to a spin-1 particle, and the antiferromagnetic inter-cell coupling $J_\mathrm{inter} > 0$ enforces the typical singlet bonds of the AKLT state between neighboring unit cells.
This is also shown in \cref{fig:ssh-aklt}~b), where the strong couplings (represented by thick lines) are now inside the unit cell.

With this in mind, we are no longer surprised that applying a homogenous microwave field to the AKLT model does not yield a gapped preparation path, since for the antiferromagnetic SSH model a homogenous microwave field also was not sufficient.
Inspired by the staggered microwave field for the SSH chain, one can also apply a staggered microwave field $\Omega \sum_i (-1)^i S^x_i$ to the AKLT model, where $S^x_i$ are now spin-1 operators.
This indeed yields a gapped preparation path from a product state of spin-1 eigenstates of the $S^x_i$ operator to the AKLT state.
Finally, since the AKLT Hamiltonian is $SO(3)$ invariant, we can replace the staggered microwave field $\pm S^x$ by a staggered, effective magnetic field $\mu \sum_i (-1)^i S^z_i$.
As shown in \cref{sec:exp-proposal}, this staggered, effective magnetic field does also provide a gapped preparation path to the ground state of Haldane phase for our Rydberg model.

\bibliography{manuscript} 

\begin{thebibliography}{54}%
\makeatletter
\providecommand \@ifxundefined [1]{%
 \@ifx{#1\undefined}
}%
\providecommand \@ifnum [1]{%
 \ifnum #1\expandafter \@firstoftwo
 \else \expandafter \@secondoftwo
 \fi
}%
\providecommand \@ifx [1]{%
 \ifx #1\expandafter \@firstoftwo
 \else \expandafter \@secondoftwo
 \fi
}%
\providecommand \natexlab [1]{#1}%
\providecommand \enquote  [1]{``#1''}%
\providecommand \bibnamefont  [1]{#1}%
\providecommand \bibfnamefont [1]{#1}%
\providecommand \citenamefont [1]{#1}%
\providecommand \href@noop [0]{\@secondoftwo}%
\providecommand \href [0]{\begingroup \@sanitize@url \@href}%
\providecommand \@href[1]{\@@startlink{#1}\@@href}%
\providecommand \@@href[1]{\endgroup#1\@@endlink}%
\providecommand \@sanitize@url [0]{\catcode `\\12\catcode `\$12\catcode
  `\&12\catcode `\#12\catcode `\^12\catcode `\_12\catcode `\%12\relax}%
\providecommand \@@startlink[1]{}%
\providecommand \@@endlink[0]{}%
\providecommand \url  [0]{\begingroup\@sanitize@url \@url }%
\providecommand \@url [1]{\endgroup\@href {#1}{\urlprefix }}%
\providecommand \urlprefix  [0]{URL }%
\providecommand \Eprint [0]{\href }%
\providecommand \doibase [0]{https://doi.org/}%
\providecommand \selectlanguage [0]{\@gobble}%
\providecommand \bibinfo  [0]{\@secondoftwo}%
\providecommand \bibfield  [0]{\@secondoftwo}%
\providecommand \translation [1]{[#1]}%
\providecommand \BibitemOpen [0]{}%
\providecommand \bibitemStop [0]{}%
\providecommand \bibitemNoStop [0]{.\EOS\space}%
\providecommand \EOS [0]{\spacefactor3000\relax}%
\providecommand \BibitemShut  [1]{\csname bibitem#1\endcsname}%
\let\auto@bib@innerbib\@empty
\bibitem [{\citenamefont {Wen}(2017)}]{wenZooQuantumtopologicalPhases2017}%
  \BibitemOpen
  \bibfield  {author} {\bibinfo {author} {\bibfnamefont {X.-G.}\ \bibnamefont
  {Wen}},\ }\bibfield  {title} {\bibinfo {title} {Zoo of quantum-topological
  phases of matter},\ }\href {https://doi.org/10.1103/RevModPhys.89.041004}
  {\bibfield  {journal} {\bibinfo  {journal} {Reviews of Modern Physics}\
  }\textbf {\bibinfo {volume} {89}},\ \bibinfo {pages} {041004} (\bibinfo
  {year} {2017})}\BibitemShut {NoStop}%
\bibitem [{\citenamefont {Stormer}\ \emph {et~al.}(1999)\citenamefont
  {Stormer}, \citenamefont {Tsui},\ and\ \citenamefont
  {Gossard}}]{stormerFractionalQuantumHall1999}%
  \BibitemOpen
  \bibfield  {author} {\bibinfo {author} {\bibfnamefont {H.~L.}\ \bibnamefont
  {Stormer}}, \bibinfo {author} {\bibfnamefont {D.~C.}\ \bibnamefont {Tsui}},\
  and\ \bibinfo {author} {\bibfnamefont {A.~C.}\ \bibnamefont {Gossard}},\
  }\bibfield  {title} {\bibinfo {title} {The fractional quantum {{Hall}}
  effect},\ }\href {https://doi.org/10.1103/RevModPhys.71.S298} {\bibfield
  {journal} {\bibinfo  {journal} {Reviews of Modern Physics}\ }\textbf
  {\bibinfo {volume} {71}},\ \bibinfo {pages} {S298} (\bibinfo {year}
  {1999})}\BibitemShut {NoStop}%
\bibitem [{\citenamefont {Verresen}\ \emph {et~al.}(2021)\citenamefont
  {Verresen}, \citenamefont {Lukin},\ and\ \citenamefont
  {Vishwanath}}]{verresenPredictionToricCode2021a}%
  \BibitemOpen
  \bibfield  {author} {\bibinfo {author} {\bibfnamefont {R.}~\bibnamefont
  {Verresen}}, \bibinfo {author} {\bibfnamefont {M.~D.}\ \bibnamefont
  {Lukin}},\ and\ \bibinfo {author} {\bibfnamefont {A.}~\bibnamefont
  {Vishwanath}},\ }\bibfield  {title} {\bibinfo {title} {Prediction of {{Toric
  Code Topological Order}} from {{Rydberg Blockade}}},\ }\href
  {https://doi.org/10.1103/PhysRevX.11.031005} {\bibfield  {journal} {\bibinfo
  {journal} {Physical Review X}\ }\textbf {\bibinfo {volume} {11}},\ \bibinfo
  {pages} {031005} (\bibinfo {year} {2021})}\BibitemShut {NoStop}%
\bibitem [{\citenamefont {Semeghini}\ \emph {et~al.}(2021)\citenamefont
  {Semeghini}, \citenamefont {Levine}, \citenamefont {Keesling}, \citenamefont
  {Ebadi}, \citenamefont {Wang}, \citenamefont {Bluvstein}, \citenamefont
  {Verresen}, \citenamefont {Pichler}, \citenamefont {Kalinowski},
  \citenamefont {Samajdar}, \citenamefont {Omran}, \citenamefont {Sachdev},
  \citenamefont {Vishwanath}, \citenamefont {Greiner}, \citenamefont
  {Vuletic},\ and\ \citenamefont
  {Lukin}}]{semeghiniProbingTopologicalSpin2021}%
  \BibitemOpen
  \bibfield  {author} {\bibinfo {author} {\bibfnamefont {G.}~\bibnamefont
  {Semeghini}}, \bibinfo {author} {\bibfnamefont {H.}~\bibnamefont {Levine}},
  \bibinfo {author} {\bibfnamefont {A.}~\bibnamefont {Keesling}}, \bibinfo
  {author} {\bibfnamefont {S.}~\bibnamefont {Ebadi}}, \bibinfo {author}
  {\bibfnamefont {T.~T.}\ \bibnamefont {Wang}}, \bibinfo {author}
  {\bibfnamefont {D.}~\bibnamefont {Bluvstein}}, \bibinfo {author}
  {\bibfnamefont {R.}~\bibnamefont {Verresen}}, \bibinfo {author}
  {\bibfnamefont {H.}~\bibnamefont {Pichler}}, \bibinfo {author} {\bibfnamefont
  {M.}~\bibnamefont {Kalinowski}}, \bibinfo {author} {\bibfnamefont
  {R.}~\bibnamefont {Samajdar}}, \bibinfo {author} {\bibfnamefont
  {A.}~\bibnamefont {Omran}}, \bibinfo {author} {\bibfnamefont
  {S.}~\bibnamefont {Sachdev}}, \bibinfo {author} {\bibfnamefont
  {A.}~\bibnamefont {Vishwanath}}, \bibinfo {author} {\bibfnamefont
  {M.}~\bibnamefont {Greiner}}, \bibinfo {author} {\bibfnamefont
  {V.}~\bibnamefont {Vuletic}},\ and\ \bibinfo {author} {\bibfnamefont {M.~D.}\
  \bibnamefont {Lukin}},\ }\bibfield  {title} {\bibinfo {title} {Probing
  {{Topological Spin Liquids}} on a {{Programmable Quantum Simulator}}},\
  }\href {https://doi.org/10.1126/science.abi8794} {\bibfield  {journal}
  {\bibinfo  {journal} {Science}\ }\textbf {\bibinfo {volume} {374}},\ \bibinfo
  {pages} {1242} (\bibinfo {year} {2021})}\BibitemShut {NoStop}%
\bibitem [{\citenamefont {Chen}\ \emph {et~al.}(2011)\citenamefont {Chen},
  \citenamefont {Gu},\ and\ \citenamefont
  {Wen}}]{chenClassificationGappedSymmetric2011}%
  \BibitemOpen
  \bibfield  {author} {\bibinfo {author} {\bibfnamefont {X.}~\bibnamefont
  {Chen}}, \bibinfo {author} {\bibfnamefont {Z.-C.}\ \bibnamefont {Gu}},\ and\
  \bibinfo {author} {\bibfnamefont {X.-G.}\ \bibnamefont {Wen}},\ }\bibfield
  {title} {\bibinfo {title} {Classification of gapped symmetric phases in
  one-dimensional spin systems},\ }\href
  {https://doi.org/10.1103/PhysRevB.83.035107} {\bibfield  {journal} {\bibinfo
  {journal} {Physical Review B}\ }\textbf {\bibinfo {volume} {83}},\ \bibinfo
  {pages} {035107} (\bibinfo {year} {2011})}\BibitemShut {NoStop}%
\bibitem [{\citenamefont
  {Haldane}(1983{\natexlab{a}})}]{haldaneNonlinearFieldTheory1983}%
  \BibitemOpen
  \bibfield  {author} {\bibinfo {author} {\bibfnamefont {F.~D.~M.}\
  \bibnamefont {Haldane}},\ }\bibfield  {title} {\bibinfo {title} {Nonlinear
  {{Field Theory}} of {{Large-Spin Heisenberg Antiferromagnets}}:
  {{Semiclassically Quantized Solitons}} of the {{One-Dimensional Easy-Axis
  N{\'e}el State}}},\ }\href {https://doi.org/10.1103/PhysRevLett.50.1153}
  {\bibfield  {journal} {\bibinfo  {journal} {Physical Review Letters}\
  }\textbf {\bibinfo {volume} {50}},\ \bibinfo {pages} {1153} (\bibinfo {year}
  {1983}{\natexlab{a}})}\BibitemShut {NoStop}%
\bibitem [{\citenamefont
  {Haldane}(1983{\natexlab{b}})}]{haldaneContinuumDynamics1Heisenberg1983}%
  \BibitemOpen
  \bibfield  {author} {\bibinfo {author} {\bibfnamefont {F.~D.~M.}\
  \bibnamefont {Haldane}},\ }\bibfield  {title} {\bibinfo {title} {Continuum
  dynamics of the 1-{{D Heisenberg}} antiferromagnet: {{Identification}} with
  the {{O}}(3) nonlinear sigma model},\ }\href
  {https://doi.org/10.1016/0375-9601(83)90631-X} {\bibfield  {journal}
  {\bibinfo  {journal} {Physics Letters A}\ }\textbf {\bibinfo {volume} {93}},\
  \bibinfo {pages} {464} (\bibinfo {year} {1983}{\natexlab{b}})}\BibitemShut
  {NoStop}%
\bibitem [{\citenamefont {Affleck}\ \emph {et~al.}(1987)\citenamefont
  {Affleck}, \citenamefont {Kennedy}, \citenamefont {Lieb},\ and\ \citenamefont
  {Tasaki}}]{affleckRigorousResultsValencebond1987}%
  \BibitemOpen
  \bibfield  {author} {\bibinfo {author} {\bibfnamefont {I.}~\bibnamefont
  {Affleck}}, \bibinfo {author} {\bibfnamefont {T.}~\bibnamefont {Kennedy}},
  \bibinfo {author} {\bibfnamefont {E.~H.}\ \bibnamefont {Lieb}},\ and\
  \bibinfo {author} {\bibfnamefont {H.}~\bibnamefont {Tasaki}},\ }\bibfield
  {title} {\bibinfo {title} {Rigorous results on valence-bond ground states in
  antiferromagnets},\ }\href {https://doi.org/10.1103/PhysRevLett.59.799}
  {\bibfield  {journal} {\bibinfo  {journal} {Physical Review Letters}\
  }\textbf {\bibinfo {volume} {59}},\ \bibinfo {pages} {799} (\bibinfo {year}
  {1987})}\BibitemShut {NoStop}%
\bibitem [{\citenamefont {Renard}\ \emph {et~al.}(2001)\citenamefont {Renard},
  \citenamefont {Regnault},\ and\ \citenamefont
  {Verdaguer}}]{renardHaldaneQuantumSpin2001}%
  \BibitemOpen
  \bibfield  {author} {\bibinfo {author} {\bibfnamefont {J.-P.}\ \bibnamefont
  {Renard}}, \bibinfo {author} {\bibfnamefont {L.-P.}\ \bibnamefont
  {Regnault}},\ and\ \bibinfo {author} {\bibfnamefont {M.}~\bibnamefont
  {Verdaguer}},\ }\bibfield  {title} {\bibinfo {title} {Haldane {{Quantum Spin
  Chains}}},\ }in\ \href {https://doi.org/10.1002/3527600841.ch2} {\emph
  {\bibinfo {booktitle} {Magnetism: {{Molecules}} to {{Materials I}}}}}\
  (\bibinfo  {publisher} {John Wiley \& Sons, Ltd},\ \bibinfo {year} {2001})\
  Chap.~\bibinfo {chapter} {2}, pp.\ \bibinfo {pages} {49--93}\BibitemShut
  {NoStop}%
\bibitem [{\citenamefont {Mishra}\ \emph {et~al.}(2021)\citenamefont {Mishra},
  \citenamefont {Catarina}, \citenamefont {Wu}, \citenamefont {Ortiz},
  \citenamefont {Jacob}, \citenamefont {Eimre}, \citenamefont {Ma},
  \citenamefont {Pignedoli}, \citenamefont {Feng}, \citenamefont {Ruffieux},
  \citenamefont {{Fern{\'a}ndez-Rossier}},\ and\ \citenamefont
  {Fasel}}]{mishraObservationFractionalEdge2021}%
  \BibitemOpen
  \bibfield  {author} {\bibinfo {author} {\bibfnamefont {S.}~\bibnamefont
  {Mishra}}, \bibinfo {author} {\bibfnamefont {G.}~\bibnamefont {Catarina}},
  \bibinfo {author} {\bibfnamefont {F.}~\bibnamefont {Wu}}, \bibinfo {author}
  {\bibfnamefont {R.}~\bibnamefont {Ortiz}}, \bibinfo {author} {\bibfnamefont
  {D.}~\bibnamefont {Jacob}}, \bibinfo {author} {\bibfnamefont
  {K.}~\bibnamefont {Eimre}}, \bibinfo {author} {\bibfnamefont
  {J.}~\bibnamefont {Ma}}, \bibinfo {author} {\bibfnamefont {C.~A.}\
  \bibnamefont {Pignedoli}}, \bibinfo {author} {\bibfnamefont {X.}~\bibnamefont
  {Feng}}, \bibinfo {author} {\bibfnamefont {P.}~\bibnamefont {Ruffieux}},
  \bibinfo {author} {\bibfnamefont {J.}~\bibnamefont
  {{Fern{\'a}ndez-Rossier}}},\ and\ \bibinfo {author} {\bibfnamefont
  {R.}~\bibnamefont {Fasel}},\ }\bibfield  {title} {\bibinfo {title}
  {Observation of fractional edge excitations in nanographene spin chains},\
  }\href {https://doi.org/10.1038/s41586-021-03842-3} {\bibfield  {journal}
  {\bibinfo  {journal} {Nature}\ }\textbf {\bibinfo {volume} {598}},\ \bibinfo
  {pages} {287} (\bibinfo {year} {2021})}\BibitemShut {NoStop}%
\bibitem [{\citenamefont {Edmunds}\ \emph {et~al.}(2024)\citenamefont
  {Edmunds}, \citenamefont {Rico}, \citenamefont {Arrazola}, \citenamefont
  {Brennen}, \citenamefont {Meth}, \citenamefont {Blatt},\ and\ \citenamefont
  {Ringbauer}}]{edmundsConstructingSpin1Haldane2024}%
  \BibitemOpen
  \bibfield  {author} {\bibinfo {author} {\bibfnamefont {C.~L.}\ \bibnamefont
  {Edmunds}}, \bibinfo {author} {\bibfnamefont {E.}~\bibnamefont {Rico}},
  \bibinfo {author} {\bibfnamefont {I.}~\bibnamefont {Arrazola}}, \bibinfo
  {author} {\bibfnamefont {G.~K.}\ \bibnamefont {Brennen}}, \bibinfo {author}
  {\bibfnamefont {M.}~\bibnamefont {Meth}}, \bibinfo {author} {\bibfnamefont
  {R.}~\bibnamefont {Blatt}},\ and\ \bibinfo {author} {\bibfnamefont
  {M.}~\bibnamefont {Ringbauer}},\ }\href@noop {} {\bibinfo {title}
  {Constructing the spin-1 {{Haldane}} phase on a qudit quantum processor}}
  (\bibinfo {year} {2024}),\ \Eprint {https://arxiv.org/abs/2408.04702}
  {arXiv:2408.04702 [quant-ph]} \BibitemShut {NoStop}%
\bibitem [{\citenamefont {Altman}\ \emph {et~al.}(2021)\citenamefont {Altman},
  \citenamefont {Brown}, \citenamefont {Carleo}, \citenamefont {Carr},
  \citenamefont {Demler}, \citenamefont {Chin}, \citenamefont {DeMarco},
  \citenamefont {Economou}, \citenamefont {Eriksson}, \citenamefont {Fu},
  \citenamefont {Greiner}, \citenamefont {Hazzard}, \citenamefont {Hulet},
  \citenamefont {Koll{\'a}r}, \citenamefont {Lev}, \citenamefont {Lukin},
  \citenamefont {Ma}, \citenamefont {Mi}, \citenamefont {Misra}, \citenamefont
  {Monroe}, \citenamefont {Murch}, \citenamefont {Nazario}, \citenamefont {Ni},
  \citenamefont {Potter}, \citenamefont {Roushan}, \citenamefont {Saffman},
  \citenamefont {{Schleier-Smith}}, \citenamefont {Siddiqi}, \citenamefont
  {Simmonds}, \citenamefont {Singh}, \citenamefont {Spielman}, \citenamefont
  {Temme}, \citenamefont {Weiss}, \citenamefont {Vu{\v c}kovi{\'c}},
  \citenamefont {Vuleti{\'c}}, \citenamefont {Ye},\ and\ \citenamefont
  {Zwierlein}}]{altmanQuantumSimulatorsArchitectures2021}%
  \BibitemOpen
  \bibfield  {author} {\bibinfo {author} {\bibfnamefont {E.}~\bibnamefont
  {Altman}}, \bibinfo {author} {\bibfnamefont {K.~R.}\ \bibnamefont {Brown}},
  \bibinfo {author} {\bibfnamefont {G.}~\bibnamefont {Carleo}}, \bibinfo
  {author} {\bibfnamefont {L.~D.}\ \bibnamefont {Carr}}, \bibinfo {author}
  {\bibfnamefont {E.}~\bibnamefont {Demler}}, \bibinfo {author} {\bibfnamefont
  {C.}~\bibnamefont {Chin}}, \bibinfo {author} {\bibfnamefont {B.}~\bibnamefont
  {DeMarco}}, \bibinfo {author} {\bibfnamefont {S.~E.}\ \bibnamefont
  {Economou}}, \bibinfo {author} {\bibfnamefont {M.~A.}\ \bibnamefont
  {Eriksson}}, \bibinfo {author} {\bibfnamefont {K.-M.~C.}\ \bibnamefont {Fu}},
  \bibinfo {author} {\bibfnamefont {M.}~\bibnamefont {Greiner}}, \bibinfo
  {author} {\bibfnamefont {K.~R.}\ \bibnamefont {Hazzard}}, \bibinfo {author}
  {\bibfnamefont {R.~G.}\ \bibnamefont {Hulet}}, \bibinfo {author}
  {\bibfnamefont {A.~J.}\ \bibnamefont {Koll{\'a}r}}, \bibinfo {author}
  {\bibfnamefont {B.~L.}\ \bibnamefont {Lev}}, \bibinfo {author} {\bibfnamefont
  {M.~D.}\ \bibnamefont {Lukin}}, \bibinfo {author} {\bibfnamefont
  {R.}~\bibnamefont {Ma}}, \bibinfo {author} {\bibfnamefont {X.}~\bibnamefont
  {Mi}}, \bibinfo {author} {\bibfnamefont {S.}~\bibnamefont {Misra}}, \bibinfo
  {author} {\bibfnamefont {C.}~\bibnamefont {Monroe}}, \bibinfo {author}
  {\bibfnamefont {K.}~\bibnamefont {Murch}}, \bibinfo {author} {\bibfnamefont
  {Z.}~\bibnamefont {Nazario}}, \bibinfo {author} {\bibfnamefont {K.-K.}\
  \bibnamefont {Ni}}, \bibinfo {author} {\bibfnamefont {A.~C.}\ \bibnamefont
  {Potter}}, \bibinfo {author} {\bibfnamefont {P.}~\bibnamefont {Roushan}},
  \bibinfo {author} {\bibfnamefont {M.}~\bibnamefont {Saffman}}, \bibinfo
  {author} {\bibfnamefont {M.}~\bibnamefont {{Schleier-Smith}}}, \bibinfo
  {author} {\bibfnamefont {I.}~\bibnamefont {Siddiqi}}, \bibinfo {author}
  {\bibfnamefont {R.}~\bibnamefont {Simmonds}}, \bibinfo {author}
  {\bibfnamefont {M.}~\bibnamefont {Singh}}, \bibinfo {author} {\bibfnamefont
  {I.}~\bibnamefont {Spielman}}, \bibinfo {author} {\bibfnamefont
  {K.}~\bibnamefont {Temme}}, \bibinfo {author} {\bibfnamefont {D.~S.}\
  \bibnamefont {Weiss}}, \bibinfo {author} {\bibfnamefont {J.}~\bibnamefont
  {Vu{\v c}kovi{\'c}}}, \bibinfo {author} {\bibfnamefont {V.}~\bibnamefont
  {Vuleti{\'c}}}, \bibinfo {author} {\bibfnamefont {J.}~\bibnamefont {Ye}},\
  and\ \bibinfo {author} {\bibfnamefont {M.}~\bibnamefont {Zwierlein}},\
  }\bibfield  {title} {\bibinfo {title} {Quantum {{Simulators}}:
  {{Architectures}} and {{Opportunities}}},\ }\href
  {https://doi.org/10.1103/PRXQuantum.2.017003} {\bibfield  {journal} {\bibinfo
   {journal} {PRX Quantum}\ }\textbf {\bibinfo {volume} {2}},\ \bibinfo {pages}
  {017003} (\bibinfo {year} {2021})}\BibitemShut {NoStop}%
\bibitem [{\citenamefont {Gross}\ and\ \citenamefont
  {Bloch}(2017)}]{grossQuantumSimulationsUltracold2017}%
  \BibitemOpen
  \bibfield  {author} {\bibinfo {author} {\bibfnamefont {C.}~\bibnamefont
  {Gross}}\ and\ \bibinfo {author} {\bibfnamefont {I.}~\bibnamefont {Bloch}},\
  }\bibfield  {title} {\bibinfo {title} {Quantum simulations with ultracold
  atoms in optical lattices},\ }\href {https://doi.org/10.1126/science.aal3837}
  {\bibfield  {journal} {\bibinfo  {journal} {Science}\ }\textbf {\bibinfo
  {volume} {357}},\ \bibinfo {pages} {995} (\bibinfo {year}
  {2017})}\BibitemShut {NoStop}%
\bibitem [{\citenamefont {Monroe}\ \emph {et~al.}(2021)\citenamefont {Monroe},
  \citenamefont {Campbell}, \citenamefont {Duan}, \citenamefont {Gong},
  \citenamefont {Gorshkov}, \citenamefont {Hess}, \citenamefont {Islam},
  \citenamefont {Kim}, \citenamefont {Linke}, \citenamefont {Pagano},
  \citenamefont {Richerme}, \citenamefont {Senko},\ and\ \citenamefont
  {Yao}}]{monroeProgrammableQuantumSimulations2021}%
  \BibitemOpen
  \bibfield  {author} {\bibinfo {author} {\bibfnamefont {C.}~\bibnamefont
  {Monroe}}, \bibinfo {author} {\bibfnamefont {W.~C.}\ \bibnamefont
  {Campbell}}, \bibinfo {author} {\bibfnamefont {L.-M.}\ \bibnamefont {Duan}},
  \bibinfo {author} {\bibfnamefont {Z.-X.}\ \bibnamefont {Gong}}, \bibinfo
  {author} {\bibfnamefont {A.~V.}\ \bibnamefont {Gorshkov}}, \bibinfo {author}
  {\bibfnamefont {P.~W.}\ \bibnamefont {Hess}}, \bibinfo {author}
  {\bibfnamefont {R.}~\bibnamefont {Islam}}, \bibinfo {author} {\bibfnamefont
  {K.}~\bibnamefont {Kim}}, \bibinfo {author} {\bibfnamefont {N.~M.}\
  \bibnamefont {Linke}}, \bibinfo {author} {\bibfnamefont {G.}~\bibnamefont
  {Pagano}}, \bibinfo {author} {\bibfnamefont {P.}~\bibnamefont {Richerme}},
  \bibinfo {author} {\bibfnamefont {C.}~\bibnamefont {Senko}},\ and\ \bibinfo
  {author} {\bibfnamefont {N.~Y.}\ \bibnamefont {Yao}},\ }\bibfield  {title}
  {\bibinfo {title} {Programmable quantum simulations of spin systems with
  trapped ions},\ }\href {https://doi.org/10.1103/RevModPhys.93.025001}
  {\bibfield  {journal} {\bibinfo  {journal} {Reviews of Modern Physics}\
  }\textbf {\bibinfo {volume} {93}},\ \bibinfo {pages} {025001} (\bibinfo
  {year} {2021})}\BibitemShut {NoStop}%
\bibitem [{\citenamefont {Dalla~Torre}\ \emph {et~al.}(2006)\citenamefont
  {Dalla~Torre}, \citenamefont {Berg},\ and\ \citenamefont
  {Altman}}]{dallatorreHiddenOrder1D2006}%
  \BibitemOpen
  \bibfield  {author} {\bibinfo {author} {\bibfnamefont {E.~G.}\ \bibnamefont
  {Dalla~Torre}}, \bibinfo {author} {\bibfnamefont {E.}~\bibnamefont {Berg}},\
  and\ \bibinfo {author} {\bibfnamefont {E.}~\bibnamefont {Altman}},\
  }\bibfield  {title} {\bibinfo {title} {Hidden {{Order}} in {{1D Bose
  Insulators}}},\ }\href {https://doi.org/10.1103/PhysRevLett.97.260401}
  {\bibfield  {journal} {\bibinfo  {journal} {Physical Review Letters}\
  }\textbf {\bibinfo {volume} {97}},\ \bibinfo {pages} {260401} (\bibinfo
  {year} {2006})}\BibitemShut {NoStop}%
\bibitem [{\citenamefont {Brennen}\ \emph {et~al.}(2007)\citenamefont
  {Brennen}, \citenamefont {Micheli},\ and\ \citenamefont
  {Zoller}}]{brennenDesigningSpin1Lattice2007}%
  \BibitemOpen
  \bibfield  {author} {\bibinfo {author} {\bibfnamefont {G.~K.}\ \bibnamefont
  {Brennen}}, \bibinfo {author} {\bibfnamefont {A.}~\bibnamefont {Micheli}},\
  and\ \bibinfo {author} {\bibfnamefont {P.}~\bibnamefont {Zoller}},\
  }\bibfield  {title} {\bibinfo {title} {Designing spin-1 lattice models using
  polar molecules},\ }\href {https://doi.org/10.1088/1367-2630/9/5/138}
  {\bibfield  {journal} {\bibinfo  {journal} {New Journal of Physics}\ }\textbf
  {\bibinfo {volume} {9}},\ \bibinfo {pages} {138} (\bibinfo {year}
  {2007})}\BibitemShut {NoStop}%
\bibitem [{\citenamefont {Cohen}\ and\ \citenamefont
  {Retzker}(2014)}]{cohenProposalVerificationHaldane2014}%
  \BibitemOpen
  \bibfield  {author} {\bibinfo {author} {\bibfnamefont {I.}~\bibnamefont
  {Cohen}}\ and\ \bibinfo {author} {\bibfnamefont {A.}~\bibnamefont
  {Retzker}},\ }\bibfield  {title} {\bibinfo {title} {Proposal for
  {{Verification}} of the {{Haldane Phase Using Trapped Ions}}},\ }\href
  {https://doi.org/10.1103/PhysRevLett.112.040503} {\bibfield  {journal}
  {\bibinfo  {journal} {Physical Review Letters}\ }\textbf {\bibinfo {volume}
  {112}},\ \bibinfo {pages} {040503} (\bibinfo {year} {2014})}\BibitemShut
  {NoStop}%
\bibitem [{\citenamefont {Senko}\ \emph {et~al.}(2015)\citenamefont {Senko},
  \citenamefont {Richerme}, \citenamefont {Smith}, \citenamefont {Lee},
  \citenamefont {Cohen}, \citenamefont {Retzker},\ and\ \citenamefont
  {Monroe}}]{senkoRealizationQuantumIntegerSpin2015}%
  \BibitemOpen
  \bibfield  {author} {\bibinfo {author} {\bibfnamefont {C.}~\bibnamefont
  {Senko}}, \bibinfo {author} {\bibfnamefont {P.}~\bibnamefont {Richerme}},
  \bibinfo {author} {\bibfnamefont {J.}~\bibnamefont {Smith}}, \bibinfo
  {author} {\bibfnamefont {A.}~\bibnamefont {Lee}}, \bibinfo {author}
  {\bibfnamefont {I.}~\bibnamefont {Cohen}}, \bibinfo {author} {\bibfnamefont
  {A.}~\bibnamefont {Retzker}},\ and\ \bibinfo {author} {\bibfnamefont
  {C.}~\bibnamefont {Monroe}},\ }\bibfield  {title} {\bibinfo {title}
  {Realization of a {{Quantum Integer-Spin Chain}} with {{Controllable
  Interactions}}},\ }\href {https://doi.org/10.1103/PhysRevX.5.021026}
  {\bibfield  {journal} {\bibinfo  {journal} {Physical Review X}\ }\textbf
  {\bibinfo {volume} {5}},\ \bibinfo {pages} {021026} (\bibinfo {year}
  {2015})}\BibitemShut {NoStop}%
\bibitem [{\citenamefont {Baran}\ and\ \citenamefont
  {Paaske}(2024)}]{baranSpin1HaldaneChains2024}%
  \BibitemOpen
  \bibfield  {author} {\bibinfo {author} {\bibfnamefont {V.~V.}\ \bibnamefont
  {Baran}}\ and\ \bibinfo {author} {\bibfnamefont {J.}~\bibnamefont {Paaske}},\
  }\bibfield  {title} {\bibinfo {title} {Spin-1 {{Haldane}} chains of
  superconductor-semiconductor hybrids},\ }\href
  {https://doi.org/10.1103/PhysRevB.110.064503} {\bibfield  {journal} {\bibinfo
   {journal} {Physical Review B}\ }\textbf {\bibinfo {volume} {110}},\ \bibinfo
  {pages} {064503} (\bibinfo {year} {2024})}\BibitemShut {NoStop}%
\bibitem [{\citenamefont {Browaeys}\ and\ \citenamefont
  {Lahaye}(2020)}]{browaeysManyBodyPhysicsIndividuallyControlled2020}%
  \BibitemOpen
  \bibfield  {author} {\bibinfo {author} {\bibfnamefont {A.}~\bibnamefont
  {Browaeys}}\ and\ \bibinfo {author} {\bibfnamefont {T.}~\bibnamefont
  {Lahaye}},\ }\bibfield  {title} {\bibinfo {title} {Many-{{Body Physics}} with
  {{Individually-Controlled Rydberg Atoms}}},\ }\href
  {https://doi.org/10.1038/s41567-019-0733-z} {\bibfield  {journal} {\bibinfo
  {journal} {Nature Physics}\ }\textbf {\bibinfo {volume} {16}},\ \bibinfo
  {pages} {132} (\bibinfo {year} {2020})}\BibitemShut {NoStop}%
\bibitem [{\citenamefont {Weber}\ \emph {et~al.}(2018)\citenamefont {Weber},
  \citenamefont {{de L{\'e}s{\'e}leuc}}, \citenamefont {Lienhard},
  \citenamefont {Barredo}, \citenamefont {Lahaye}, \citenamefont {Browaeys},\
  and\ \citenamefont {B{\"u}chler}}]{weberTopologicallyProtectedEdge2018}%
  \BibitemOpen
  \bibfield  {author} {\bibinfo {author} {\bibfnamefont {S.}~\bibnamefont
  {Weber}}, \bibinfo {author} {\bibfnamefont {S.}~\bibnamefont {{de
  L{\'e}s{\'e}leuc}}}, \bibinfo {author} {\bibfnamefont {V.}~\bibnamefont
  {Lienhard}}, \bibinfo {author} {\bibfnamefont {D.}~\bibnamefont {Barredo}},
  \bibinfo {author} {\bibfnamefont {T.}~\bibnamefont {Lahaye}}, \bibinfo
  {author} {\bibfnamefont {A.}~\bibnamefont {Browaeys}},\ and\ \bibinfo
  {author} {\bibfnamefont {H.~P.}\ \bibnamefont {B{\"u}chler}},\ }\bibfield
  {title} {\bibinfo {title} {Topologically protected edge states in small
  {{Rydberg}} systems},\ }\href {https://doi.org/10.1088/2058-9565/aaca47}
  {\bibfield  {journal} {\bibinfo  {journal} {Quantum Science and Technology}\
  }\textbf {\bibinfo {volume} {3}},\ \bibinfo {pages} {044001} (\bibinfo {year}
  {2018})}\BibitemShut {NoStop}%
\bibitem [{\citenamefont {De~L{\'e}s{\'e}leuc}\ \emph
  {et~al.}(2019)\citenamefont {De~L{\'e}s{\'e}leuc}, \citenamefont {Lienhard},
  \citenamefont {Scholl}, \citenamefont {Barredo}, \citenamefont {Weber},
  \citenamefont {Lang}, \citenamefont {B{\"u}chler}, \citenamefont {Lahaye},\
  and\ \citenamefont
  {Browaeys}}]{deleseleucObservationSymmetryprotectedTopological2019}%
  \BibitemOpen
  \bibfield  {author} {\bibinfo {author} {\bibfnamefont {S.}~\bibnamefont
  {De~L{\'e}s{\'e}leuc}}, \bibinfo {author} {\bibfnamefont {V.}~\bibnamefont
  {Lienhard}}, \bibinfo {author} {\bibfnamefont {P.}~\bibnamefont {Scholl}},
  \bibinfo {author} {\bibfnamefont {D.}~\bibnamefont {Barredo}}, \bibinfo
  {author} {\bibfnamefont {S.}~\bibnamefont {Weber}}, \bibinfo {author}
  {\bibfnamefont {N.}~\bibnamefont {Lang}}, \bibinfo {author} {\bibfnamefont
  {H.~P.}\ \bibnamefont {B{\"u}chler}}, \bibinfo {author} {\bibfnamefont
  {T.}~\bibnamefont {Lahaye}},\ and\ \bibinfo {author} {\bibfnamefont
  {A.}~\bibnamefont {Browaeys}},\ }\bibfield  {title} {\bibinfo {title}
  {Observation of a symmetry-protected topological phase of interacting bosons
  with {{Rydberg}} atoms},\ }\href {https://doi.org/10.1126/science.aav9105}
  {\bibfield  {journal} {\bibinfo  {journal} {Science}\ }\textbf {\bibinfo
  {volume} {365}},\ \bibinfo {pages} {775} (\bibinfo {year}
  {2019})}\BibitemShut {NoStop}%
\bibitem [{\citenamefont {Lienhard}\ \emph {et~al.}(2020)\citenamefont
  {Lienhard}, \citenamefont {Scholl}, \citenamefont {Weber}, \citenamefont
  {Barredo}, \citenamefont {{de L{\'e}s{\'e}leuc}}, \citenamefont {Bai},
  \citenamefont {Lang}, \citenamefont {Fleischhauer}, \citenamefont
  {B{\"u}chler}, \citenamefont {Lahaye},\ and\ \citenamefont
  {Browaeys}}]{lienhardRealizationDensitydependentPeierls2020}%
  \BibitemOpen
  \bibfield  {author} {\bibinfo {author} {\bibfnamefont {V.}~\bibnamefont
  {Lienhard}}, \bibinfo {author} {\bibfnamefont {P.}~\bibnamefont {Scholl}},
  \bibinfo {author} {\bibfnamefont {S.}~\bibnamefont {Weber}}, \bibinfo
  {author} {\bibfnamefont {D.}~\bibnamefont {Barredo}}, \bibinfo {author}
  {\bibfnamefont {S.}~\bibnamefont {{de L{\'e}s{\'e}leuc}}}, \bibinfo {author}
  {\bibfnamefont {R.}~\bibnamefont {Bai}}, \bibinfo {author} {\bibfnamefont
  {N.}~\bibnamefont {Lang}}, \bibinfo {author} {\bibfnamefont {M.}~\bibnamefont
  {Fleischhauer}}, \bibinfo {author} {\bibfnamefont {H.~P.}\ \bibnamefont
  {B{\"u}chler}}, \bibinfo {author} {\bibfnamefont {T.}~\bibnamefont
  {Lahaye}},\ and\ \bibinfo {author} {\bibfnamefont {A.}~\bibnamefont
  {Browaeys}},\ }\bibfield  {title} {\bibinfo {title} {Realization of a
  density-dependent {{Peierls}} phase in a synthetic, spin-orbit coupled
  {{Rydberg}} system},\ }\href {https://doi.org/10.1103/PhysRevX.10.021031}
  {\bibfield  {journal} {\bibinfo  {journal} {Physical Review X}\ }\textbf
  {\bibinfo {volume} {10}},\ \bibinfo {pages} {021031} (\bibinfo {year}
  {2020})}\BibitemShut {NoStop}%
\bibitem [{\citenamefont {S{\o}rensen}\ \emph {et~al.}(2005)\citenamefont
  {S{\o}rensen}, \citenamefont {Demler},\ and\ \citenamefont
  {Lukin}}]{sorensenFractionalQuantumHall2005}%
  \BibitemOpen
  \bibfield  {author} {\bibinfo {author} {\bibfnamefont {A.~S.}\ \bibnamefont
  {S{\o}rensen}}, \bibinfo {author} {\bibfnamefont {E.}~\bibnamefont
  {Demler}},\ and\ \bibinfo {author} {\bibfnamefont {M.~D.}\ \bibnamefont
  {Lukin}},\ }\bibfield  {title} {\bibinfo {title} {Fractional {{Quantum Hall
  States}} of {{Atoms}} in {{Optical Lattices}}},\ }\href
  {https://doi.org/10.1103/PhysRevLett.94.086803} {\bibfield  {journal}
  {\bibinfo  {journal} {Physical Review Letters}\ }\textbf {\bibinfo {volume}
  {94}},\ \bibinfo {pages} {086803} (\bibinfo {year} {2005})}\BibitemShut
  {NoStop}%
\bibitem [{\citenamefont {B{\"u}chler}\ \emph {et~al.}(2005)\citenamefont
  {B{\"u}chler}, \citenamefont {Hermele}, \citenamefont {Huber}, \citenamefont
  {Fisher},\ and\ \citenamefont {Zoller}}]{buchlerAtomicQuantumSimulator2005}%
  \BibitemOpen
  \bibfield  {author} {\bibinfo {author} {\bibfnamefont {H.~P.}\ \bibnamefont
  {B{\"u}chler}}, \bibinfo {author} {\bibfnamefont {M.}~\bibnamefont
  {Hermele}}, \bibinfo {author} {\bibfnamefont {S.~D.}\ \bibnamefont {Huber}},
  \bibinfo {author} {\bibfnamefont {M.~P.~A.}\ \bibnamefont {Fisher}},\ and\
  \bibinfo {author} {\bibfnamefont {P.}~\bibnamefont {Zoller}},\ }\bibfield
  {title} {\bibinfo {title} {Atomic {{Quantum Simulator}} for {{Lattice Gauge
  Theories}} and {{Ring Exchange Models}}},\ }\href
  {https://doi.org/10.1103/PhysRevLett.95.040402} {\bibfield  {journal}
  {\bibinfo  {journal} {Physical Review Letters}\ }\textbf {\bibinfo {volume}
  {95}},\ \bibinfo {pages} {040402} (\bibinfo {year} {2005})}\BibitemShut
  {NoStop}%
\bibitem [{\citenamefont {Duan}\ \emph {et~al.}(2003)\citenamefont {Duan},
  \citenamefont {Demler},\ and\ \citenamefont
  {Lukin}}]{duanControllingSpinExchange2003}%
  \BibitemOpen
  \bibfield  {author} {\bibinfo {author} {\bibfnamefont {L.-M.}\ \bibnamefont
  {Duan}}, \bibinfo {author} {\bibfnamefont {E.}~\bibnamefont {Demler}},\ and\
  \bibinfo {author} {\bibfnamefont {M.~D.}\ \bibnamefont {Lukin}},\ }\bibfield
  {title} {\bibinfo {title} {Controlling {{Spin Exchange Interactions}} of
  {{Ultracold Atoms}} in {{Optical Lattices}}},\ }\href
  {https://doi.org/10.1103/PhysRevLett.91.090402} {\bibfield  {journal}
  {\bibinfo  {journal} {Physical Review Letters}\ }\textbf {\bibinfo {volume}
  {91}},\ \bibinfo {pages} {090402} (\bibinfo {year} {2003})}\BibitemShut
  {NoStop}%
\bibitem [{\citenamefont {Miroshnychenko}\ \emph {et~al.}(2006)\citenamefont
  {Miroshnychenko}, \citenamefont {Alt}, \citenamefont {Dotsenko},
  \citenamefont {F{\"o}rster}, \citenamefont {Khudaverdyan}, \citenamefont
  {Rauschenbeutel},\ and\ \citenamefont
  {Meschede}}]{miroshnychenkoPrecisionPreparationStrings2006}%
  \BibitemOpen
  \bibfield  {author} {\bibinfo {author} {\bibfnamefont {Y.}~\bibnamefont
  {Miroshnychenko}}, \bibinfo {author} {\bibfnamefont {W.}~\bibnamefont {Alt}},
  \bibinfo {author} {\bibfnamefont {I.}~\bibnamefont {Dotsenko}}, \bibinfo
  {author} {\bibfnamefont {L.}~\bibnamefont {F{\"o}rster}}, \bibinfo {author}
  {\bibfnamefont {M.}~\bibnamefont {Khudaverdyan}}, \bibinfo {author}
  {\bibfnamefont {A.}~\bibnamefont {Rauschenbeutel}},\ and\ \bibinfo {author}
  {\bibfnamefont {D.}~\bibnamefont {Meschede}},\ }\bibfield  {title} {\bibinfo
  {title} {Precision preparation of strings of trapped neutral atoms},\ }\href
  {https://doi.org/10.1088/1367-2630/8/9/191} {\bibfield  {journal} {\bibinfo
  {journal} {New Journal of Physics}\ }\textbf {\bibinfo {volume} {8}},\
  \bibinfo {pages} {191} (\bibinfo {year} {2006})}\BibitemShut {NoStop}%
\bibitem [{\citenamefont {Kim}\ \emph {et~al.}(2016)\citenamefont {Kim},
  \citenamefont {Lee}, \citenamefont {Lee}, \citenamefont {Jo}, \citenamefont
  {Song},\ and\ \citenamefont {Ahn}}]{kimSituSingleatomArray2016}%
  \BibitemOpen
  \bibfield  {author} {\bibinfo {author} {\bibfnamefont {H.}~\bibnamefont
  {Kim}}, \bibinfo {author} {\bibfnamefont {W.}~\bibnamefont {Lee}}, \bibinfo
  {author} {\bibfnamefont {H.-g.}\ \bibnamefont {Lee}}, \bibinfo {author}
  {\bibfnamefont {H.}~\bibnamefont {Jo}}, \bibinfo {author} {\bibfnamefont
  {Y.}~\bibnamefont {Song}},\ and\ \bibinfo {author} {\bibfnamefont
  {J.}~\bibnamefont {Ahn}},\ }\bibfield  {title} {\bibinfo {title} {In situ
  single-atom array synthesis using dynamic holographic optical tweezers},\
  }\href {https://doi.org/10.1038/ncomms13317} {\bibfield  {journal} {\bibinfo
  {journal} {Nature Communications}\ }\textbf {\bibinfo {volume} {7}},\
  \bibinfo {pages} {13317} (\bibinfo {year} {2016})}\BibitemShut {NoStop}%
\bibitem [{\citenamefont {Endres}\ \emph {et~al.}(2016)\citenamefont {Endres},
  \citenamefont {Bernien}, \citenamefont {Keesling}, \citenamefont {Levine},
  \citenamefont {Anschuetz}, \citenamefont {Krajenbrink}, \citenamefont
  {Senko}, \citenamefont {Vuletic}, \citenamefont {Greiner},\ and\
  \citenamefont {Lukin}}]{endresAtomatomAssemblyDefectfree2016}%
  \BibitemOpen
  \bibfield  {author} {\bibinfo {author} {\bibfnamefont {M.}~\bibnamefont
  {Endres}}, \bibinfo {author} {\bibfnamefont {H.}~\bibnamefont {Bernien}},
  \bibinfo {author} {\bibfnamefont {A.}~\bibnamefont {Keesling}}, \bibinfo
  {author} {\bibfnamefont {H.}~\bibnamefont {Levine}}, \bibinfo {author}
  {\bibfnamefont {E.~R.}\ \bibnamefont {Anschuetz}}, \bibinfo {author}
  {\bibfnamefont {A.}~\bibnamefont {Krajenbrink}}, \bibinfo {author}
  {\bibfnamefont {C.}~\bibnamefont {Senko}}, \bibinfo {author} {\bibfnamefont
  {V.}~\bibnamefont {Vuletic}}, \bibinfo {author} {\bibfnamefont
  {M.}~\bibnamefont {Greiner}},\ and\ \bibinfo {author} {\bibfnamefont {M.~D.}\
  \bibnamefont {Lukin}},\ }\bibfield  {title} {\bibinfo {title} {Atom-by-atom
  assembly of defect-free one-dimensional cold atom arrays},\ }\href
  {https://doi.org/10.1126/science.aah3752} {\bibfield  {journal} {\bibinfo
  {journal} {Science}\ }\textbf {\bibinfo {volume} {354}},\ \bibinfo {pages}
  {1024} (\bibinfo {year} {2016})}\BibitemShut {NoStop}%
\bibitem [{\citenamefont {Barredo}\ \emph {et~al.}(2016)\citenamefont
  {Barredo}, \citenamefont {{de L{\'e}s{\'e}leuc}}, \citenamefont {Lienhard},
  \citenamefont {Lahaye},\ and\ \citenamefont
  {Browaeys}}]{barredoAtomatomAssemblerDefectfree2016}%
  \BibitemOpen
  \bibfield  {author} {\bibinfo {author} {\bibfnamefont {D.}~\bibnamefont
  {Barredo}}, \bibinfo {author} {\bibfnamefont {S.}~\bibnamefont {{de
  L{\'e}s{\'e}leuc}}}, \bibinfo {author} {\bibfnamefont {V.}~\bibnamefont
  {Lienhard}}, \bibinfo {author} {\bibfnamefont {T.}~\bibnamefont {Lahaye}},\
  and\ \bibinfo {author} {\bibfnamefont {A.}~\bibnamefont {Browaeys}},\
  }\bibfield  {title} {\bibinfo {title} {An atom-by-atom assembler of
  defect-free arbitrary two-dimensional atomic arrays},\ }\href
  {https://doi.org/10.1126/science.aah3778} {\bibfield  {journal} {\bibinfo
  {journal} {Science}\ }\textbf {\bibinfo {volume} {354}},\ \bibinfo {pages}
  {1021} (\bibinfo {year} {2016})}\BibitemShut {NoStop}%
\bibitem [{\citenamefont {Labuhn}\ \emph {et~al.}(2016)\citenamefont {Labuhn},
  \citenamefont {Barredo}, \citenamefont {Ravets}, \citenamefont {{de
  L{\'e}s{\'e}leuc}}, \citenamefont {Macr{\`i}}, \citenamefont {Lahaye},\ and\
  \citenamefont {Browaeys}}]{labuhnTunableTwodimensionalArrays2016}%
  \BibitemOpen
  \bibfield  {author} {\bibinfo {author} {\bibfnamefont {H.}~\bibnamefont
  {Labuhn}}, \bibinfo {author} {\bibfnamefont {D.}~\bibnamefont {Barredo}},
  \bibinfo {author} {\bibfnamefont {S.}~\bibnamefont {Ravets}}, \bibinfo
  {author} {\bibfnamefont {S.}~\bibnamefont {{de L{\'e}s{\'e}leuc}}}, \bibinfo
  {author} {\bibfnamefont {T.}~\bibnamefont {Macr{\`i}}}, \bibinfo {author}
  {\bibfnamefont {T.}~\bibnamefont {Lahaye}},\ and\ \bibinfo {author}
  {\bibfnamefont {A.}~\bibnamefont {Browaeys}},\ }\bibfield  {title} {\bibinfo
  {title} {Tunable two-dimensional arrays of single {{Rydberg}} atoms for
  realizing quantum {{Ising}} models},\ }\href
  {https://doi.org/10.1038/nature18274} {\bibfield  {journal} {\bibinfo
  {journal} {Nature}\ }\textbf {\bibinfo {volume} {534}},\ \bibinfo {pages}
  {667} (\bibinfo {year} {2016})}\BibitemShut {NoStop}%
\bibitem [{\citenamefont {Weber}\ \emph {et~al.}(2022)\citenamefont {Weber},
  \citenamefont {Bai}, \citenamefont {Makki}, \citenamefont {M{\"o}gerle},
  \citenamefont {Lahaye}, \citenamefont {Browaeys}, \citenamefont {Daghofer},
  \citenamefont {Lang},\ and\ \citenamefont
  {B{\"u}chler}}]{weberExperimentallyAccessibleScheme2022}%
  \BibitemOpen
  \bibfield  {author} {\bibinfo {author} {\bibfnamefont {S.}~\bibnamefont
  {Weber}}, \bibinfo {author} {\bibfnamefont {R.}~\bibnamefont {Bai}}, \bibinfo
  {author} {\bibfnamefont {N.}~\bibnamefont {Makki}}, \bibinfo {author}
  {\bibfnamefont {J.}~\bibnamefont {M{\"o}gerle}}, \bibinfo {author}
  {\bibfnamefont {T.}~\bibnamefont {Lahaye}}, \bibinfo {author} {\bibfnamefont
  {A.}~\bibnamefont {Browaeys}}, \bibinfo {author} {\bibfnamefont
  {M.}~\bibnamefont {Daghofer}}, \bibinfo {author} {\bibfnamefont
  {N.}~\bibnamefont {Lang}},\ and\ \bibinfo {author} {\bibfnamefont {H.~P.}\
  \bibnamefont {B{\"u}chler}},\ }\bibfield  {title} {\bibinfo {title}
  {Experimentally {{Accessible Scheme}} for a {{Fractional Chern Insulator}} in
  {{Rydberg Atoms}}},\ }\href {https://doi.org/10.1103/PRXQuantum.3.030302}
  {\bibfield  {journal} {\bibinfo  {journal} {PRX Quantum}\ }\textbf {\bibinfo
  {volume} {3}},\ \bibinfo {pages} {030302} (\bibinfo {year}
  {2022})}\BibitemShut {NoStop}%
\bibitem [{\citenamefont {Safinya}\ \emph {et~al.}(1981)\citenamefont
  {Safinya}, \citenamefont {Delpech}, \citenamefont {Gounand}, \citenamefont
  {Sandner},\ and\ \citenamefont
  {Gallagher}}]{safinyaResonantRydbergAtomRydbergAtomCollisions1981}%
  \BibitemOpen
  \bibfield  {author} {\bibinfo {author} {\bibfnamefont {K.~A.}\ \bibnamefont
  {Safinya}}, \bibinfo {author} {\bibfnamefont {J.~F.}\ \bibnamefont
  {Delpech}}, \bibinfo {author} {\bibfnamefont {F.}~\bibnamefont {Gounand}},
  \bibinfo {author} {\bibfnamefont {W.}~\bibnamefont {Sandner}},\ and\ \bibinfo
  {author} {\bibfnamefont {T.~F.}\ \bibnamefont {Gallagher}},\ }\bibfield
  {title} {\bibinfo {title} {Resonant {{Rydberg-Atom-Rydberg-Atom
  Collisions}}},\ }\href {https://doi.org/10.1103/PhysRevLett.47.405}
  {\bibfield  {journal} {\bibinfo  {journal} {Physical Review Letters}\
  }\textbf {\bibinfo {volume} {47}},\ \bibinfo {pages} {405} (\bibinfo {year}
  {1981})}\BibitemShut {NoStop}%
\bibitem [{\citenamefont {Ryabtsev}\ \emph {et~al.}(2010)\citenamefont
  {Ryabtsev}, \citenamefont {Tretyakov}, \citenamefont {Beterov},\ and\
  \citenamefont {Entin}}]{ryabtsevObservationStarkTunedForster2010}%
  \BibitemOpen
  \bibfield  {author} {\bibinfo {author} {\bibfnamefont {I.~I.}\ \bibnamefont
  {Ryabtsev}}, \bibinfo {author} {\bibfnamefont {D.~B.}\ \bibnamefont
  {Tretyakov}}, \bibinfo {author} {\bibfnamefont {I.~I.}\ \bibnamefont
  {Beterov}},\ and\ \bibinfo {author} {\bibfnamefont {V.~M.}\ \bibnamefont
  {Entin}},\ }\bibfield  {title} {\bibinfo {title} {Observation of the
  {{Stark-Tuned F{\"o}rster Resonance}} between {{Two Rydberg Atoms}}},\ }\href
  {https://doi.org/10.1103/PhysRevLett.104.073003} {\bibfield  {journal}
  {\bibinfo  {journal} {Physical Review Letters}\ }\textbf {\bibinfo {volume}
  {104}},\ \bibinfo {pages} {073003} (\bibinfo {year} {2010})}\BibitemShut
  {NoStop}%
\bibitem [{\citenamefont {M{\"o}gerle}(2022)}]{MScJohannesMoegerle}%
  \BibitemOpen
  \bibfield  {author} {\bibinfo {author} {\bibfnamefont {J.}~\bibnamefont
  {M{\"o}gerle}},\ }\href
  {https://www.itp3.uni-stuttgart.de/downloads/theses/MasterThesis_JohannesMoegerle_2022.pdf}
  {\bibinfo {title} {Spin 1 haldane phase in one-dimensional systems of rydberg
  atoms}} (\bibinfo {year} {2022}),\ \bibinfo {note} {{{Master thesis}},
  {{University of Stuttgart}}}\BibitemShut {NoStop}%
\bibitem [{\citenamefont {Liu}\ \emph {et~al.}(2024)\citenamefont {Liu},
  \citenamefont {Bintz}, \citenamefont {Block}, \citenamefont {Samajdar},
  \citenamefont {Kemp},\ and\ \citenamefont
  {Yao}}]{liuSupersoliditySimplexPhases2024a}%
  \BibitemOpen
  \bibfield  {author} {\bibinfo {author} {\bibfnamefont {V.~S.}\ \bibnamefont
  {Liu}}, \bibinfo {author} {\bibfnamefont {M.}~\bibnamefont {Bintz}}, \bibinfo
  {author} {\bibfnamefont {M.}~\bibnamefont {Block}}, \bibinfo {author}
  {\bibfnamefont {R.}~\bibnamefont {Samajdar}}, \bibinfo {author}
  {\bibfnamefont {J.}~\bibnamefont {Kemp}},\ and\ \bibinfo {author}
  {\bibfnamefont {N.~Y.}\ \bibnamefont {Yao}},\ }\href
  {https://doi.org/10.48550/arXiv.2407.17554} {\bibinfo {title} {Supersolidity
  and {{Simplex Phases}} in {{Spin-1 Rydberg Atom Arrays}}}} (\bibinfo {year}
  {2024})\BibitemShut {NoStop}%
\bibitem [{\citenamefont {Homeier}\ \emph {et~al.}(2024)\citenamefont
  {Homeier}, \citenamefont {Harris}, \citenamefont {Blatz}, \citenamefont
  {Geier}, \citenamefont {Hollerith}, \citenamefont {Schollw{\"o}ck},
  \citenamefont {Grusdt},\ and\ \citenamefont
  {Bohrdt}}]{homeierAntiferromagneticBosonic2024}%
  \BibitemOpen
  \bibfield  {author} {\bibinfo {author} {\bibfnamefont {L.}~\bibnamefont
  {Homeier}}, \bibinfo {author} {\bibfnamefont {T.~J.}\ \bibnamefont {Harris}},
  \bibinfo {author} {\bibfnamefont {T.}~\bibnamefont {Blatz}}, \bibinfo
  {author} {\bibfnamefont {S.}~\bibnamefont {Geier}}, \bibinfo {author}
  {\bibfnamefont {S.}~\bibnamefont {Hollerith}}, \bibinfo {author}
  {\bibfnamefont {U.}~\bibnamefont {Schollw{\"o}ck}}, \bibinfo {author}
  {\bibfnamefont {F.}~\bibnamefont {Grusdt}},\ and\ \bibinfo {author}
  {\bibfnamefont {A.}~\bibnamefont {Bohrdt}},\ }\bibfield  {title} {\bibinfo
  {title} {Antiferromagnetic bosonic $t$-${{J}}$ models and their quantum
  simulation in tweezer arrays},\ }\href
  {https://doi.org/10.1103/PhysRevLett.132.230401} {\bibfield  {journal}
  {\bibinfo  {journal} {Physical Review Letters}\ }\textbf {\bibinfo {volume}
  {132}},\ \bibinfo {pages} {230401} (\bibinfo {year} {2024})}\BibitemShut
  {NoStop}%
\bibitem [{\citenamefont {Sompet}\ \emph {et~al.}(2022)\citenamefont {Sompet},
  \citenamefont {Hirthe}, \citenamefont {Bourgund}, \citenamefont {Chalopin},
  \citenamefont {Bibo}, \citenamefont {Koepsell}, \citenamefont {Bojovi{\'c}},
  \citenamefont {Verresen}, \citenamefont {Pollmann}, \citenamefont {Salomon},
  \citenamefont {Gross}, \citenamefont {Hilker},\ and\ \citenamefont
  {Bloch}}]{sompetRealisingSymmetryProtectedHaldane2022}%
  \BibitemOpen
  \bibfield  {author} {\bibinfo {author} {\bibfnamefont {P.}~\bibnamefont
  {Sompet}}, \bibinfo {author} {\bibfnamefont {S.}~\bibnamefont {Hirthe}},
  \bibinfo {author} {\bibfnamefont {D.}~\bibnamefont {Bourgund}}, \bibinfo
  {author} {\bibfnamefont {T.}~\bibnamefont {Chalopin}}, \bibinfo {author}
  {\bibfnamefont {J.}~\bibnamefont {Bibo}}, \bibinfo {author} {\bibfnamefont
  {J.}~\bibnamefont {Koepsell}}, \bibinfo {author} {\bibfnamefont
  {P.}~\bibnamefont {Bojovi{\'c}}}, \bibinfo {author} {\bibfnamefont
  {R.}~\bibnamefont {Verresen}}, \bibinfo {author} {\bibfnamefont
  {F.}~\bibnamefont {Pollmann}}, \bibinfo {author} {\bibfnamefont
  {G.}~\bibnamefont {Salomon}}, \bibinfo {author} {\bibfnamefont
  {C.}~\bibnamefont {Gross}}, \bibinfo {author} {\bibfnamefont {T.~A.}\
  \bibnamefont {Hilker}},\ and\ \bibinfo {author} {\bibfnamefont
  {I.}~\bibnamefont {Bloch}},\ }\bibfield  {title} {\bibinfo {title} {Realising
  the {{Symmetry-Protected Haldane Phase}} in {{Fermi-Hubbard Ladders}}},\
  }\href {https://doi.org/10.1038/s41586-022-04688-z} {\bibfield  {journal}
  {\bibinfo  {journal} {Nature}\ }\textbf {\bibinfo {volume} {606}},\ \bibinfo
  {pages} {484} (\bibinfo {year} {2022})}\BibitemShut {NoStop}%
\bibitem [{\citenamefont {Weber}\ \emph {et~al.}(2017)\citenamefont {Weber},
  \citenamefont {Tresp}, \citenamefont {Menke}, \citenamefont {Urvoy},
  \citenamefont {Firstenberg}, \citenamefont {B{\"u}chler},\ and\ \citenamefont
  {Hofferberth}}]{weberTutorialCalculationRydberg2017}%
  \BibitemOpen
  \bibfield  {author} {\bibinfo {author} {\bibfnamefont {S.}~\bibnamefont
  {Weber}}, \bibinfo {author} {\bibfnamefont {C.}~\bibnamefont {Tresp}},
  \bibinfo {author} {\bibfnamefont {H.}~\bibnamefont {Menke}}, \bibinfo
  {author} {\bibfnamefont {A.}~\bibnamefont {Urvoy}}, \bibinfo {author}
  {\bibfnamefont {O.}~\bibnamefont {Firstenberg}}, \bibinfo {author}
  {\bibfnamefont {H.~P.}\ \bibnamefont {B{\"u}chler}},\ and\ \bibinfo {author}
  {\bibfnamefont {S.}~\bibnamefont {Hofferberth}},\ }\bibfield  {title}
  {\bibinfo {title} {Tutorial: {{Calculation}} of {{Rydberg}} interaction
  potentials},\ }\href {https://doi.org/10.1088/1361-6455/aa743a} {\bibfield
  {journal} {\bibinfo  {journal} {Journal of Physics B: Atomic, Molecular and
  Optical Physics}\ }\textbf {\bibinfo {volume} {50}},\ \bibinfo {pages}
  {133001} (\bibinfo {year} {2017})}\BibitemShut {NoStop}%
\bibitem [{\citenamefont {Pollmann}\ \emph {et~al.}(2010)\citenamefont
  {Pollmann}, \citenamefont {Berg}, \citenamefont {Turner},\ and\ \citenamefont
  {Oshikawa}}]{pollmannEntanglementSpectrumTopological2010}%
  \BibitemOpen
  \bibfield  {author} {\bibinfo {author} {\bibfnamefont {F.}~\bibnamefont
  {Pollmann}}, \bibinfo {author} {\bibfnamefont {E.}~\bibnamefont {Berg}},
  \bibinfo {author} {\bibfnamefont {A.~M.}\ \bibnamefont {Turner}},\ and\
  \bibinfo {author} {\bibfnamefont {M.}~\bibnamefont {Oshikawa}},\ }\bibfield
  {title} {\bibinfo {title} {Entanglement spectrum of a topological phase in
  one dimension},\ }\href {https://doi.org/10.1103/PhysRevB.81.064439}
  {\bibfield  {journal} {\bibinfo  {journal} {Physical Review B}\ }\textbf
  {\bibinfo {volume} {81}},\ \bibinfo {pages} {064439} (\bibinfo {year}
  {2010})}\BibitemShut {NoStop}%
\bibitem [{\citenamefont {Pollmann}\ and\ \citenamefont
  {Turner}(2012)}]{pollmannDetectionSymmetryProtected2012}%
  \BibitemOpen
  \bibfield  {author} {\bibinfo {author} {\bibfnamefont {F.}~\bibnamefont
  {Pollmann}}\ and\ \bibinfo {author} {\bibfnamefont {A.~M.}\ \bibnamefont
  {Turner}},\ }\bibfield  {title} {\bibinfo {title} {Detection of {{Symmetry
  Protected Topological Phases}} in {{1D}}},\ }\href
  {https://doi.org/10.1103/PhysRevB.86.125441} {\bibfield  {journal} {\bibinfo
  {journal} {Physical Review B}\ }\textbf {\bibinfo {volume} {86}},\ \bibinfo
  {pages} {125441} (\bibinfo {year} {2012})}\BibitemShut {NoStop}%
\bibitem [{\citenamefont {Chen}\ \emph {et~al.}(2003)\citenamefont {Chen},
  \citenamefont {Hida},\ and\ \citenamefont
  {Sanctuary}}]{chenGroundstatePhaseDiagram2003}%
  \BibitemOpen
  \bibfield  {author} {\bibinfo {author} {\bibfnamefont {W.}~\bibnamefont
  {Chen}}, \bibinfo {author} {\bibfnamefont {K.}~\bibnamefont {Hida}},\ and\
  \bibinfo {author} {\bibfnamefont {B.~C.}\ \bibnamefont {Sanctuary}},\
  }\bibfield  {title} {\bibinfo {title} {Ground-state phase diagram of {{S}} =
  1 {{XXZ}} chains with uniaxial single-ion-type anisotropy},\ }\href
  {https://doi.org/10.1103/PhysRevB.67.104401} {\bibfield  {journal} {\bibinfo
  {journal} {Physical Review B}\ }\textbf {\bibinfo {volume} {67}},\ \bibinfo
  {pages} {104401} (\bibinfo {year} {2003})}\BibitemShut {NoStop}%
\bibitem [{\citenamefont {Gong}\ \emph {et~al.}(2016)\citenamefont {Gong},
  \citenamefont {Maghrebi}, \citenamefont {Hu}, \citenamefont {{Foss-Feig}},
  \citenamefont {Richerme}, \citenamefont {Monroe},\ and\ \citenamefont
  {Gorshkov}}]{gongKaleidoscopeQuantumPhases2016}%
  \BibitemOpen
  \bibfield  {author} {\bibinfo {author} {\bibfnamefont {Z.-X.}\ \bibnamefont
  {Gong}}, \bibinfo {author} {\bibfnamefont {M.~F.}\ \bibnamefont {Maghrebi}},
  \bibinfo {author} {\bibfnamefont {A.}~\bibnamefont {Hu}}, \bibinfo {author}
  {\bibfnamefont {M.}~\bibnamefont {{Foss-Feig}}}, \bibinfo {author}
  {\bibfnamefont {P.}~\bibnamefont {Richerme}}, \bibinfo {author}
  {\bibfnamefont {C.}~\bibnamefont {Monroe}},\ and\ \bibinfo {author}
  {\bibfnamefont {A.~V.}\ \bibnamefont {Gorshkov}},\ }\bibfield  {title}
  {\bibinfo {title} {Kaleidoscope of quantum phases in a long-range interacting
  spin-1 chain},\ }\href {https://doi.org/10.1103/PhysRevB.93.205115}
  {\bibfield  {journal} {\bibinfo  {journal} {Physical Review B}\ }\textbf
  {\bibinfo {volume} {93}},\ \bibinfo {pages} {205115} (\bibinfo {year}
  {2016})}\BibitemShut {NoStop}%
\bibitem [{\citenamefont {Hauschild}\ and\ \citenamefont
  {Pollmann}(2018)}]{hauschildEfficientNumericalSimulations2018}%
  \BibitemOpen
  \bibfield  {author} {\bibinfo {author} {\bibfnamefont {J.}~\bibnamefont
  {Hauschild}}\ and\ \bibinfo {author} {\bibfnamefont {F.}~\bibnamefont
  {Pollmann}},\ }\bibfield  {title} {\bibinfo {title} {Efficient numerical
  simulations with {{Tensor Networks}}: {{Tensor Network Python}}
  ({{TeNPy}})},\ }\href {https://doi.org/10.21468/SciPostPhysLectNotes.5}
  {\bibfield  {journal} {\bibinfo  {journal} {SciPost Physics Lecture Notes}\
  ,\ \bibinfo {pages} {5}} (\bibinfo {year} {2018})}\BibitemShut {NoStop}%
\bibitem [{\citenamefont {Peter}\ \emph {et~al.}(2012)\citenamefont {Peter},
  \citenamefont {M{\"u}ller}, \citenamefont {Wessel},\ and\ \citenamefont
  {B{\"u}chler}}]{peterAnomalousBehaviorSpin2012}%
  \BibitemOpen
  \bibfield  {author} {\bibinfo {author} {\bibfnamefont {D.}~\bibnamefont
  {Peter}}, \bibinfo {author} {\bibfnamefont {S.}~\bibnamefont {M{\"u}ller}},
  \bibinfo {author} {\bibfnamefont {S.}~\bibnamefont {Wessel}},\ and\ \bibinfo
  {author} {\bibfnamefont {H.~P.}\ \bibnamefont {B{\"u}chler}},\ }\bibfield
  {title} {\bibinfo {title} {Anomalous {{Behavior}} of {{Spin Systems}} with
  {{Dipolar Interactions}}},\ }\href
  {https://doi.org/10.1103/PhysRevLett.109.025303} {\bibfield  {journal}
  {\bibinfo  {journal} {Physical Review Letters}\ }\textbf {\bibinfo {volume}
  {109}},\ \bibinfo {pages} {025303} (\bibinfo {year} {2012})}\BibitemShut
  {NoStop}%
\bibitem [{\citenamefont {Schulz}(1986)}]{schulzPhaseDiagramsCorrelation1986}%
  \BibitemOpen
  \bibfield  {author} {\bibinfo {author} {\bibfnamefont {H.~J.}\ \bibnamefont
  {Schulz}},\ }\bibfield  {title} {\bibinfo {title} {Phase diagrams and
  correlation exponents for quantum spin chains of arbitrary spin quantum
  number},\ }\href {https://doi.org/10.1103/PhysRevB.34.6372} {\bibfield
  {journal} {\bibinfo  {journal} {Physical Review B}\ }\textbf {\bibinfo
  {volume} {34}},\ \bibinfo {pages} {6372} (\bibinfo {year}
  {1986})}\BibitemShut {NoStop}%
\bibitem [{\citenamefont {Kitazawa}\ \emph {et~al.}(1996)\citenamefont
  {Kitazawa}, \citenamefont {Nomura},\ and\ \citenamefont
  {Okamoto}}]{kitazawaPhaseDiagram11996}%
  \BibitemOpen
  \bibfield  {author} {\bibinfo {author} {\bibfnamefont {A.}~\bibnamefont
  {Kitazawa}}, \bibinfo {author} {\bibfnamefont {K.}~\bibnamefont {Nomura}},\
  and\ \bibinfo {author} {\bibfnamefont {K.}~\bibnamefont {Okamoto}},\
  }\bibfield  {title} {\bibinfo {title} {Phase {{Diagram}} of {{S}} = 1
  {{Bond-Alternating}} {{{\emph{XXZ}}}} {{Chains}}},\ }\href
  {https://doi.org/10.1103/PhysRevLett.76.4038} {\bibfield  {journal} {\bibinfo
   {journal} {Physical Review Letters}\ }\textbf {\bibinfo {volume} {76}},\
  \bibinfo {pages} {4038} (\bibinfo {year} {1996})}\BibitemShut {NoStop}%
\bibitem [{\citenamefont {Heng~Su}\ \emph {et~al.}(2012)\citenamefont
  {Heng~Su}, \citenamefont {Young~Cho}, \citenamefont {Li}, \citenamefont
  {Wang},\ and\ \citenamefont {Zhou}}]{hengsuNonlocalCorrelationsHaldane2012}%
  \BibitemOpen
  \bibfield  {author} {\bibinfo {author} {\bibfnamefont {Y.}~\bibnamefont
  {Heng~Su}}, \bibinfo {author} {\bibfnamefont {S.}~\bibnamefont {Young~Cho}},
  \bibinfo {author} {\bibfnamefont {B.}~\bibnamefont {Li}}, \bibinfo {author}
  {\bibfnamefont {H.-L.}\ \bibnamefont {Wang}},\ and\ \bibinfo {author}
  {\bibfnamefont {H.-Q.}\ \bibnamefont {Zhou}},\ }\bibfield  {title} {\bibinfo
  {title} {Non-local {{Correlations}} in the {{Haldane Phase}} for an {{XXZ
  Spin-1 Chain}}: {{A Perspective}} from {{Infinite Matrix Product State
  Representation}}},\ }\href {https://doi.org/10.1143/JPSJ.81.074003}
  {\bibfield  {journal} {\bibinfo  {journal} {Journal of the Physical Society
  of Japan}\ }\textbf {\bibinfo {volume} {81}},\ \bibinfo {pages} {074003}
  (\bibinfo {year} {2012})}\BibitemShut {NoStop}%
\bibitem [{\citenamefont {Kitazawa}\ and\ \citenamefont
  {Nomura}(1997)}]{kitazawaCriticalProperties11997}%
  \BibitemOpen
  \bibfield  {author} {\bibinfo {author} {\bibfnamefont {A.}~\bibnamefont
  {Kitazawa}}\ and\ \bibinfo {author} {\bibfnamefont {K.}~\bibnamefont
  {Nomura}},\ }\bibfield  {title} {\bibinfo {title} {Critical {{Properties}} of
  {{S}} = 1 {{Bond-Alternating XXZ Chains}} and {{Hidden}} ${Z}_2 \times {Z}_2$
  {{Symmetry}}},\ }\href {https://doi.org/10.1143/JPSJ.66.3944} {\bibfield
  {journal} {\bibinfo  {journal} {Journal of the Physical Society of Japan}\
  }\textbf {\bibinfo {volume} {66}},\ \bibinfo {pages} {3944} (\bibinfo {year}
  {1997})}\BibitemShut {NoStop}%
\bibitem [{\citenamefont {Zaletel}\ \emph {et~al.}(2015)\citenamefont
  {Zaletel}, \citenamefont {Mong}, \citenamefont {Karrasch}, \citenamefont
  {Moore},\ and\ \citenamefont
  {Pollmann}}]{zaletelTimeevolvingMatrixProduct2015}%
  \BibitemOpen
  \bibfield  {author} {\bibinfo {author} {\bibfnamefont {M.~P.}\ \bibnamefont
  {Zaletel}}, \bibinfo {author} {\bibfnamefont {R.~S.~K.}\ \bibnamefont
  {Mong}}, \bibinfo {author} {\bibfnamefont {C.}~\bibnamefont {Karrasch}},
  \bibinfo {author} {\bibfnamefont {J.~E.}\ \bibnamefont {Moore}},\ and\
  \bibinfo {author} {\bibfnamefont {F.}~\bibnamefont {Pollmann}},\ }\bibfield
  {title} {\bibinfo {title} {Time-evolving a matrix product state with
  long-ranged interactions},\ }\href
  {https://doi.org/10.1103/PhysRevB.91.165112} {\bibfield  {journal} {\bibinfo
  {journal} {Physical Review B}\ }\textbf {\bibinfo {volume} {91}},\ \bibinfo
  {pages} {165112} (\bibinfo {year} {2015})}\BibitemShut {NoStop}%
\bibitem [{\citenamefont {{Garc{\'i}a-Ripoll}}\ \emph
  {et~al.}(2004)\citenamefont {{Garc{\'i}a-Ripoll}}, \citenamefont
  {{Martin-Delgado}},\ and\ \citenamefont
  {Cirac}}]{garcia-ripollImplementationSpinHamiltonians2004}%
  \BibitemOpen
  \bibfield  {author} {\bibinfo {author} {\bibfnamefont {J.~J.}\ \bibnamefont
  {{Garc{\'i}a-Ripoll}}}, \bibinfo {author} {\bibfnamefont {M.~A.}\
  \bibnamefont {{Martin-Delgado}}},\ and\ \bibinfo {author} {\bibfnamefont
  {J.~I.}\ \bibnamefont {Cirac}},\ }\bibfield  {title} {\bibinfo {title}
  {Implementation of {{Spin Hamiltonians}} in {{Optical Lattices}}},\ }\href
  {https://doi.org/10.1103/PhysRevLett.93.250405} {\bibfield  {journal}
  {\bibinfo  {journal} {Physical Review Letters}\ }\textbf {\bibinfo {volume}
  {93}},\ \bibinfo {pages} {250405} (\bibinfo {year} {2004})}\BibitemShut
  {NoStop}%
\bibitem [{\citenamefont {Chen}\ \emph {et~al.}(2023)\citenamefont {Chen},
  \citenamefont {Bornet}, \citenamefont {Bintz}, \citenamefont {Emperauger},
  \citenamefont {Leclerc}, \citenamefont {Liu}, \citenamefont {Scholl},
  \citenamefont {Barredo}, \citenamefont {Hauschild}, \citenamefont
  {Chatterjee}, \citenamefont {Schuler}, \citenamefont {L{\"a}uchli},
  \citenamefont {Zaletel}, \citenamefont {Lahaye}, \citenamefont {Yao},\ and\
  \citenamefont {Browaeys}}]{chenContinuousSymmetryBreaking2023}%
  \BibitemOpen
  \bibfield  {author} {\bibinfo {author} {\bibfnamefont {C.}~\bibnamefont
  {Chen}}, \bibinfo {author} {\bibfnamefont {G.}~\bibnamefont {Bornet}},
  \bibinfo {author} {\bibfnamefont {M.}~\bibnamefont {Bintz}}, \bibinfo
  {author} {\bibfnamefont {G.}~\bibnamefont {Emperauger}}, \bibinfo {author}
  {\bibfnamefont {L.}~\bibnamefont {Leclerc}}, \bibinfo {author} {\bibfnamefont
  {V.~S.}\ \bibnamefont {Liu}}, \bibinfo {author} {\bibfnamefont
  {P.}~\bibnamefont {Scholl}}, \bibinfo {author} {\bibfnamefont
  {D.}~\bibnamefont {Barredo}}, \bibinfo {author} {\bibfnamefont
  {J.}~\bibnamefont {Hauschild}}, \bibinfo {author} {\bibfnamefont
  {S.}~\bibnamefont {Chatterjee}}, \bibinfo {author} {\bibfnamefont
  {M.}~\bibnamefont {Schuler}}, \bibinfo {author} {\bibfnamefont {A.~M.}\
  \bibnamefont {L{\"a}uchli}}, \bibinfo {author} {\bibfnamefont {M.~P.}\
  \bibnamefont {Zaletel}}, \bibinfo {author} {\bibfnamefont {T.}~\bibnamefont
  {Lahaye}}, \bibinfo {author} {\bibfnamefont {N.~Y.}\ \bibnamefont {Yao}},\
  and\ \bibinfo {author} {\bibfnamefont {A.}~\bibnamefont {Browaeys}},\
  }\bibfield  {title} {\bibinfo {title} {Continuous symmetry breaking in a
  two-dimensional {{Rydberg}} array},\ }\href
  {https://doi.org/10.1038/s41586-023-05859-2} {\bibfield  {journal} {\bibinfo
  {journal} {Nature}\ }\textbf {\bibinfo {volume} {616}},\ \bibinfo {pages}
  {691} (\bibinfo {year} {2023})}\BibitemShut {NoStop}%
\bibitem [{\citenamefont {Mögerle}\ \emph {et~al.}(2025)\citenamefont
  {Mögerle}, \citenamefont {Brechtelsbauer}, \citenamefont {Gea~Caballero},
  \citenamefont {Prior}, \citenamefont {Emperauger}, \citenamefont {Bornet},
  \citenamefont {Chen}, \citenamefont {Lahaye}, \citenamefont {Browaeys},\ and\
  \citenamefont {Büchler}}]{DARUS-4990_2025}%
  \BibitemOpen
  \bibfield  {author} {\bibinfo {author} {\bibfnamefont {J.}~\bibnamefont
  {Mögerle}}, \bibinfo {author} {\bibfnamefont {K.}~\bibnamefont
  {Brechtelsbauer}}, \bibinfo {author} {\bibfnamefont {A.~T.}\ \bibnamefont
  {Gea~Caballero}}, \bibinfo {author} {\bibfnamefont {J.}~\bibnamefont
  {Prior}}, \bibinfo {author} {\bibfnamefont {G.}~\bibnamefont {Emperauger}},
  \bibinfo {author} {\bibfnamefont {G.}~\bibnamefont {Bornet}}, \bibinfo
  {author} {\bibfnamefont {C.}~\bibnamefont {Chen}}, \bibinfo {author}
  {\bibfnamefont {T.}~\bibnamefont {Lahaye}}, \bibinfo {author} {\bibfnamefont
  {A.}~\bibnamefont {Browaeys}},\ and\ \bibinfo {author} {\bibfnamefont
  {H.~P.}\ \bibnamefont {Büchler}},\ }\href
  {https://doi.org/10.18419/DARUS-4990} {\bibinfo {title} {{Data for ``Spin-1
  Haldane phase in a chain of Rydberg atoms''}}} (\bibinfo {year}
  {2025})\BibitemShut {NoStop}%
\bibitem [{\citenamefont {Bravyi}\ \emph {et~al.}(2011)\citenamefont {Bravyi},
  \citenamefont {DiVincenzo},\ and\ \citenamefont
  {Loss}}]{bravyiSchriefferWolffTransformationQuantum2011}%
  \BibitemOpen
  \bibfield  {author} {\bibinfo {author} {\bibfnamefont {S.}~\bibnamefont
  {Bravyi}}, \bibinfo {author} {\bibfnamefont {D.}~\bibnamefont {DiVincenzo}},\
  and\ \bibinfo {author} {\bibfnamefont {D.}~\bibnamefont {Loss}},\ }\bibfield
  {title} {\bibinfo {title} {Schrieffer-{{Wolff}} transformation for quantum
  many-body systems},\ }\href {https://doi.org/10.1016/j.aop.2011.06.004}
  {\bibfield  {journal} {\bibinfo  {journal} {Annals of Physics}\ }\textbf
  {\bibinfo {volume} {326}},\ \bibinfo {pages} {2793} (\bibinfo {year}
  {2011})}\BibitemShut {NoStop}%
\end{thebibliography}%
\end{document}